\documentclass[JHEP,11pt]{article}
\usepackage{amsmath,amssymb}
\usepackage[dvipsnames]{xcolor}
\usepackage{xspace}
\usepackage{ifdraft}
\usepackage{bm}
\usepackage{epstopdf}
\usepackage{anyfontsize}
\usepackage{slashed}
\usepackage{diagbox}
\usepackage{colortbl}
\usepackage{tocloft}

\usepackage{bbold}
\usepackage{tabularx}
\usepackage{graphicx}
\usepackage{stackengine}

\usepackage{jheppub}
\usepackage{xcolor}
\hypersetup{
    colorlinks,
    linkcolor={nhpRed},
    citecolor={nhpBlue},
    urlcolor={nhpBlue}
}

\usepackage[framemethod=default]{mdframed}
\newmdenv[skipabove=7pt,
skipbelow=7pt,
rightline=false,
leftline=false,
topline=false,
bottomline=false,
backgroundcolor=gray!15,
linecolor=gray,
innerleftmargin=5pt,
innerrightmargin=5pt,
innertopmargin=5pt,
innerbottommargin=5pt,
leftmargin=0cm,
rightmargin=0cm,
linewidth=4pt]{eBox}

\usepackage{tikz-feynman} 

\tikzfeynmanset{compat=1.1.0}
\tikzfeynmanset{/tikzfeynman/warn luatex = false}

\definecolor{jaxoblue}{HTML}{0086FF}

\newcommand{\cP}{\rm{cov.}\pi}

\usepackage{xspace}
\newcommand{\Poincare}{Poincar\'e\xspace}

\def\taua{{{\rm t}}}
\def\bartaua{{{\bar {\rm t}}}}

\definecolor{nhpRed}{RGB}{161,0,0}
\definecolor{nhp4}{RGB}{203, 4, 31}
\definecolor{nhp3}{RGB}{244,99,30}
\definecolor{nhp2}{RGB}{255,159,0}
\definecolor{nhp1}{RGB}{48,152,152}
\definecolor{nhpBlue}{RGB}{0,100,144}
\definecolor{cutred}{RGB}{219,56,49}
\definecolor{hgreen}{RGB}{25,176,146}
\definecolor{hgreen1}{RGB}{175,230,175}
\definecolor{hblue}{RGB}{52,152,219}
\definecolor{hblue1}{RGB}{255,255,166}
\definecolor{hred}{RGB}{216,83,117}
\definecolor{hred1}{RGB}{255,155,155}
\definecolor{cutred}{RGB}{219,56,49}
\definecolor{hgrey4}{RGB}{75,75,75}
\definecolor{hgrey5}{RGB}{50,50,50}
\definecolor{hgrey3}{RGB}{100,100,100}
\definecolor{hgrey}{RGB}{125,125,125}
\definecolor{hgrey2}{RGB}{125,125,125}
\definecolor{hgrey1}{RGB}{150,150,150}
\definecolor{hgrey0}{RGB}{175,175,175}
\definecolor{darkgreen}{RGB}{59,126,108}

\newcommand{\threeGraph}[6]{ {
\begin{tikzpicture}[baseline=(current  bounding  box.center)]
\begin{feynman}
\vertex(mid1) at (0,0);
\vertex (a1) at (-1*1.3, -.8*1.3) {\(\,#4\,\)};
\vertex (a2) at (-1*1.3, .8*1.3) {\(\,#5\,\)};
\vertex (a3) at (1.3*1.3, 0*1.3) {\(\,#6\,\)};
\diagram{
(mid1) --[#1, thick](a1),
(mid1) --[#2, thick](a2),
(mid1) -- [#3, thick](a3),
};
\end{feynman}
\end{tikzpicture}
}
}

\newcommand{\threeTableGraph}[5]{ {
\begin{tikzpicture}[baseline=(current  bounding  box.center)]
\begin{feynman}
\vertex [blob,#5](mid1) at (0,0) {\(\,#4\,\)};
\vertex [white](mid2) at (0,0) {\(\,#4\,\)};
\vertex (a1) at (1, -.8) {};
\vertex (a2) at (1, .8) {};
\vertex (a3) at (-1.36, 0) {};
\diagram{
(mid1) --[#1, thick](a1),
(mid1) --[#2, thick](a2),
(mid1) -- [#3, thick](a3),
};
\end{feynman}
\end{tikzpicture}
}
}

\newcommand{\fourTableGraph}[6]{ {
\begin{tikzpicture}[baseline=(current  bounding  box.center)]
\begin{feynman}
\vertex [blob,#6](mid1) at (0,0) {\(\,#5\,\)};
\vertex [white](mid2) at (0,0) {\(\,#5\,\)};
\vertex (a1) at (1, -.8) {};
\vertex (a2) at (1, .8) {};
\vertex (a3) at (-1, 0.8) {};
\vertex (a4) at (-1,- 0.8) {};
\diagram{
(mid1) --[#1, thick](a1),
(mid1) --[#2, thick](a2),
(mid1) -- [#3, thick](a3),
(mid1) -- [#4, thick](a4),
};
\end{feynman}
\end{tikzpicture}
}
}

\newcommand{\fiveTableGraph}[7]{ {
\begin{tikzpicture}[baseline=(current  bounding  box.center)]
\begin{feynman}
\vertex [blob,#7](mid1) at (0,0) {\(\,#6\,\)};
\vertex [white](mid2) at (0,0) {\(\,#6\,\)};
\vertex (a1) at (1, -.8) {};
\vertex (a2) at (1, .8) {};
\vertex (a3) at (-1, 0.8) {};
\vertex (a4) at (-1,- 0.8) {};
\vertex (a5) at (0,1.4) {};
\diagram{
(mid1) --[#1, thick](a1),
(mid1) --[#2, thick](a2),
(mid1) -- [#3, thick](a3),
(mid1) -- [#4, thick](a4),
(mid1) -- [#5, thick](a5),
};
\end{feynman}
\end{tikzpicture}
}
}

\newcommand{\sixTableGraph}[8]{ {
\begin{tikzpicture}[baseline=(current  bounding  box.center)]
\begin{feynman}
\vertex [blob,#7](mid1) at (0,0) {\(\,#8\,\)};
\vertex [white](mid2) at (0,0) {\(\,#8\,\)};
\vertex (a1) at (1, -.8) {};
\vertex (a2) at (1, .8) {};
\vertex (a3) at (-1, 0.8) {};
\vertex (a4) at (-1,- 0.8) {};
\vertex (a5) at (0,1.24) {};
\vertex (a6) at (0,-1.24) {};
\diagram{
(mid1) --[#1, thick](a1),
(mid1) --[#2, thick](a2),
(mid1) -- [#3, thick](a3),
(mid1) -- [#4, thick](a4),
(mid1) -- [#5, thick](a5),
(mid1) -- [#6, thick](a6),
};
\end{feynman}
\end{tikzpicture}
}
}

\newcommand{\cutDiagramSix}{ {
\begin{tikzpicture}[baseline=(current  bounding  box.center)]
\begin{feynman}
\vertex [blob,hgrey](mid1) at (0,0) {$\Lambda^2/g$};
\vertex [white](mid2) at (0,0) {$\Lambda^2/g$};
\vertex (a3) at (-1, 0.8) {};
\vertex (a4) at (-1,- 0.8) {};
\vertex (a5) at (0,1.24) {};
\vertex (a6) at (0,-1.24) {};
\vertex [blob,hgrey](mid3) at (2,0) {$g$};
\vertex [white](mid4) at (2,0) {$g$};
\vertex (a1) at (3, -.8) {};
\vertex (a2) at (3, .8) {};
\vertex (b2) at (1.1,-0.8){};
\vertex (b1) at (1.1,0.8){};
\diagram{
(mid3) --[ thick](a1),
(mid3) --[thick](a2),
(mid3) --[gluon, thick](mid1),
(mid1) -- [thick](a3),
(mid1) -- [thick](a4),
(mid1) -- [thick](a5),
(mid1) -- [thick](a6),
(b1) -- [scalar,thick, cutred](b2)
};
\end{feynman}
\end{tikzpicture}
}
}

\newcommand{\fivegraphMostScalarIntGluePropOp}[8]{ {
\begin{tikzpicture}[baseline=(current  bounding  box.center)]
\begin{feynman}
\vertex (a1) at (-2.2, -1.2) {\(#1\)};
\vertex (a2) at (-2.6, 0) {\(#2\)};
\vertex (a5) at (-2.2, 1.2) {\(#7\)};
\vertex [blob,hgrey](mid1) at (-1,0) {\(\,#5\,\)};
\vertex [white](mid3) at (-1,0) {\(\,#5\,\)};
\vertex [blob,#8](mid2) at (0.7,0) {\(\,#6\,\)};
\vertex [white](mid4) at (0.7,0) {\(\,#6\,\)};
\vertex (a3) at (1.6, 1.2) {\(#3\)};
\vertex (a4) at (1.6, -1.2) {\(#4\)};
\vertex (a7) at (0, -0.9);
\vertex (a6) at (0, 0.9);
\diagram{
(mid2) --[ thick](a3),
(mid2) --[ thick](a4),
(mid2) -- [gluon, thick](mid1),
(a1)--[ thick](mid1),
(a2) --[ thick](mid1),
(mid1) --[gluon, thick](a5),
(a6)--[scalar, thick,cutred](a7)
};
\end{feynman}
\end{tikzpicture}
}
}

\newcommand{\fivegraphGluePropOp}[9]{ {
\begin{tikzpicture}[baseline=(current  bounding  box.center)]
\begin{feynman}
\vertex (a1) at (-2.2, -1.2) {\(#1\)};
\vertex (a2) at (-2.6, 0) {\(#2\)};
\vertex (a5) at (-2.2, 1.2) {\(#7\)};
\vertex [blob,#9](mid1) at (-1,0) {\(\,#5\,\)};
\vertex [white](mid3) at (-1,0) {\(\,#5\,\)};
\vertex [blob,#8](mid2) at (0.7,0) {\(\,#6\,\)};
\vertex [white](mid4) at (0.7,0) {\(\,#6\,\)};
\vertex (a3) at (1.6, 1.2) {\(#3\)};
\vertex (a4) at (1.6, -1.2) {\(#4\)};
\vertex (a7) at (0, -0.9);
\vertex (a6) at (0, 0.9);
\diagram{
(mid2) --[gluon, thick](a3),
(mid2) --[gluon, thick](a4),
(mid2) -- [gluon, thick](mid1),
(a1)--[gluon, thick](mid1),
(a2) -- [gluon, thick](mid1),
(mid1) --[gluon, thick](a5),
(a6)--[scalar, thick,cutred](a7)
};
\end{feynman}
\end{tikzpicture}
}
}

\newcommand{\fivegraphScalarPropOp}[9]{ {
\begin{tikzpicture}[baseline=(current  bounding  box.center)]
\begin{feynman}
\vertex (a1) at (-2.2, -1.2) {\(#1\)};
\vertex (a2) at (-2.6, 0) {\(#2\)};
\vertex (a5) at (-2.2, 1.2) {\(#7\)};
\vertex [blob,#9](mid1) at (-1,0) {\(\,#5\,\)};
\vertex [white](mid3) at (-1,0) {\(\,#5\,\)};
\vertex [blob,#8](mid2) at (0.7,0) {\(\,#6\,\)};
\vertex [white](mid4) at (0.7,0) {\(\,#6\,\)};
\vertex (a3) at (1.6, 1.2) {\(#3\)};
\vertex (a4) at (1.6, -1.2) {\(#4\)};
\vertex (a7) at (0, -0.9);
\vertex (a6) at (0, 0.9);
\diagram{
(mid2) --[gluon, thick](a3),
(mid2) --[gluon, thick](a4),
(mid2) -- [gluon, thick](mid1),
(a1)--[ thick](mid1),
(a2) -- [gluon, thick](mid1),
(mid1) --[ thick](a5),
(a6)--[scalar, thick,cutred](a7)
};
\end{feynman}
\end{tikzpicture}
}
}

\newcommand{\fivegraphMostScalarExtGluePropOp}[8]{ {
\begin{tikzpicture}[baseline=(current  bounding  box.center)]
\begin{feynman}
\vertex (a1) at (-2.2, -1.2) {\(#1\)};
\vertex (a2) at (-2.6, 0) {\(#2\)};
\vertex (a5) at (-2.2, 1.2) {\(#7\)};
\vertex [blob,hgrey](mid1) at (-1,0) {\(\,#5\,\)};
\vertex [white](mid3) at (-1,0) {\(\,#5\,\)};
\vertex [blob,#8](mid2) at (0.7,0) {\(\,#6\,\)};
\vertex [white](mid4) at (0.7,0) {\(\,#6\,\)};
\vertex (a3) at (1.6, 1.2) {\(#3\)};
\vertex (a4) at (1.6, -1.2) {\(#4\)};
\vertex (a7) at (0, -0.9);
\vertex (a6) at (0, 0.9);
\diagram{
(mid2) --[ thick](a3),
(mid2) --[gluon, thick](a4),
(mid2) --[ thick](mid1),
(a1)--[ thick](mid1),
(a2) --[ thick](mid1),
(mid1) --[ thick](a5),
(a6)--[scalar, thick,cutred](a7)
};
\end{feynman}
\end{tikzpicture}
}
}

\newcommand{\fivegraphBasis}[5]{ {
\begin{tikzpicture}[baseline=(current  bounding  box.center)]
\begin{feynman}
\vertex (a1) at (-1, -0.9) {\(#1\)};
\vertex (a2) at (-1, 0.9) {\(#2\)};
\vertex (mid1) at (-1,0);
\vertex (mid2) at (0.3,0);
\vertex (mid3) at (1.6, 0);
\vertex (a5) at (0.3, 0.9) {\(#3\)};
\vertex (a3) at (1.6, 0.9) {\(#4\)};
\vertex (a4) at (1.6, -0.9) {\(#5\)};
\diagram{
(mid2) --[ thick](a5),
(mid2) -- [ thick,scalar](mid3),
(a1)--[ thick](mid1),
(a2) -- [ thick](mid1), 
(mid1)--[ thick,scalar](mid2), 
(mid3)--[ thick](a3), 
(a4) -- [ thick,gluon](mid3)
};
\end{feynman}
\end{tikzpicture}
}
}

\newcommand{\fivegraphChild}[5]{ {
\begin{tikzpicture}[baseline=(current  bounding  box.center)]
\begin{feynman}
\vertex (a1) at (-1, -0.9) {\(#1\)};
\vertex (a2) at (-1, 0.9) {\(#2\)};
\vertex (mid1) at (-1,0);
\vertex (mid2) at (0.3,0);
\vertex (mid3) at (1.6, 0);
\vertex (a5) at (0.3, 0.9) {\(#3\)};
\vertex (a3) at (1.6, 0.9) {\(#4\)};
\vertex (a4) at (1.6, -0.9) {\(#5\)};
\diagram{
(mid2) --[ thick,gluon](a5),
(mid2) -- [ thick,scalar](mid3),
(a1)--[ thick](mid1),
(a2) -- [ thick](mid1), 
(mid1)--[ thick,scalar](mid2), 
(mid3)--[ thick](a3), 
(a4) -- [ thick](mid3)
};
\end{feynman}
\end{tikzpicture}
}
}

\newcommand{\fivegraphMcut}[5]{ {
\begin{tikzpicture}[baseline=(current  bounding  box.center)]
\begin{feynman}
\vertex (a1) at (-1, -0.8) {\(#1\)};
\vertex (a2) at (-1, 0.8) {\(#2\)};
\vertex(mid1) at (-1,0);
\vertex (midAb) at (0,0) ;
\vertex (midAt) at (0,.8) {\(#3\)};
\vertex (mid2) at (1,0);
\vertex (a3) at (1, 0.8) {\(#4\)};
\vertex (a4) at (1,- 0.8) {\(#5\)};
\vertex (a5) at (0.5, -0.7);
\vertex (a6) at (0.5, 0.7);
\vertex (a7) at (-0.5, -0.7);
\vertex (a8) at (-0.5, 0.7);
\diagram{
(midAb) --[ thick,gluon](midAt),
(mid2) --[ thick](a3),
(mid2) --[ thick](a4),
(mid2) -- [ thick,scalar](midAb),
(midAb) --[ thick,scalar](mid1),
(a1) --[ thick](mid1),
(a2) --[ thick](mid1),
(a5)--[scalar, thick,cutred](a6),
(a7)--[scalar, thick,cutred](a8)
};
\end{feynman}
\end{tikzpicture}
}
}

\newcommand{\halfLadder}[1]{ {
\begin{tikzpicture}[baseline=(current  bounding  box.center)]
\begin{feynman}
\vertex (mid1) at (-2, 0);
\vertex (mid2) at (-1, 0);
\vertex (mid3) at (1, 0);
\vertex [dot, scale=0.4](d1) at (-0.4, .5){};
\vertex [dot, scale=0.4](d2) at (0, .5){};
\vertex [dot, scale=0.4](d3) at (.4, .5){};
\vertex (a1) at (-3, 0) {$1$};
\vertex (a2) at (-2, 1) {$\sigma(2)$};
\vertex (a3) at (-1, 1) {$\sigma(3)$};
\vertex (a4) at (1, 1) {$\sigma(n-1)$};
\vertex (a5) at (2, 0) {$n$};
\diagram{
(mid1)--[thick](a1),
(mid1)--[thick](a2),
(mid1)--[thick](mid2),
(mid3)--[thick](mid2),
(mid2)--[thick](a3),
(mid3)--[thick](a4),
(mid3)--[thick](a5),
};
\end{feynman}
\end{tikzpicture}
}
}

\newcommand{\tchannel}{ {
\begin{tikzpicture}[baseline=(current  bounding  box.center)]
\begin{feynman}
\vertex (mid1) at (0, 0);
\vertex (mid2) at (0,1);
\vertex (a1) at (-1, 0) {$a_1$};
\vertex (a2) at (-.75, 1.75) {$a_2$};
\vertex (a3) at (.75, 1.75) {$a_3$};
\vertex (a4) at (1, 0) {$a_4$};
\diagram{
(mid1)--[thick](a1),
(mid1)--[thick](a4),
(mid2)--[thick](a2),
(mid2)--[thick](a3),
(mid1)--[thick,hred](mid2)
};
\end{feynman}
\end{tikzpicture}
}
}

\newcommand{\schannel}{ {
\begin{tikzpicture}[baseline=(current  bounding  box.center)]
\begin{feynman}
\vertex (mid1) at (-0.5, 0);
\vertex (mid2) at (0.5,0);
\vertex (a1) at (-1.5, 0) {$a_1$};
\vertex (a2) at (-.5, 1) {$a_2$};
\vertex (a3) at (.5, 1) {$a_3$};
\vertex (a4) at (1.5, 0) {$a_4$};
\diagram{
(mid1)--[thick](a1),
(mid1)--[thick](a2),
(mid2)--[thick](a4),
(mid2)--[thick](a3),
(mid1)--[thick,hred](mid2)
};
\end{feynman}
\end{tikzpicture}
}
}

\newcommand{\uchannel}{ {
\begin{tikzpicture}[baseline=(current  bounding  box.center)]
\begin{feynman}
\vertex (mid1) at (-0.5, 0);
\vertex (mid2) at (0.5,0);
\vertex (a1) at (-1.5, 0) {$a_1$};
\vertex (a2) at (-.5, 1) {$a_3$};
\vertex (a3) at (.5, 1) {$a_2$};
\vertex (a4) at (1.5, 0) {$a_4$};
\diagram{
(mid1)--[thick](a1),
(mid1)--[thick](a2),
(mid2)--[thick](a4),
(mid2)--[thick](a3),
(mid1)--[thick,hred](mid2)
};
\end{feynman}
\end{tikzpicture}
}
}

 \def\draftnote#1{{\color{red}\it #1}} 

\def\sect#1{section~\ref{#1}}
\def\Fig#1{fig.~{\ref{#1}}}
\def\Fig#1{Fig.~{\ref{#1}}}

\def\spa#1.#2{\left\langle#1\,#2\right\rangle}
\def\spb#1.#2{\left[#1\,#2\right]}
\def\spash#1.#2{\spa{\smash{#1}}.{\smash{#2}}}
\def\spbsh#1.#2{\spb{\smash{#1}}.{\smash{#2}}}
\def\sand#1.#2.#3{%
\left\langle\smash{#1}{\vphantom1}^{-}\right|{#2}%
\left|\smash{#3}{\vphantom1}^{-}\right\rangle}
\def\sandpp#1.#2.#3{%
\left\langle\smash{#1}{\vphantom1}^{+}\right|{#2}%
\left|\smash{#3}{\vphantom1}^{+}\right\rangle}
\def\sandpm#1.#2.#3{%
\left\langle\smash{#1}{\vphantom1}^{+}\right|{#2}%
\left|\smash{#3}{\vphantom1}^{-}\right\rangle}
\def\sandmp#1.#2.#3{%
\left\langle\smash{#1}{\vphantom1}^{-}\right|{#2}%
\left|\smash{#3}{\vphantom1}^{+}\right\rangle}

\def\Tr{\, {\rm Tr}}
\def\tr{\, {\rm tr}}

\def\nn{\nonumber}
\def\sec#1{section~\ref{#1}}
\def\eqn#1{eq.~(\ref{#1})}

\def\secref#1{section~\ref{#1}}

\def\NeqFour{{{\cal N}=4}}

\def\be{\begin{equation}}
\def\ee{\end{equation}}
\def\bea{\begin{eqnarray}}
\def\eea{\end{eqnarray}}
\def\ba{\begin{eqnarray}}
\def\ea{\end{eqnarray}}

 \definecolor{MattOrange}{rgb}{1.0,0.4,0.2}

\newcommand{\andd}{\ , \quad \text{and}  \quad}
\newcommand{\forr}{\ , \quad \text{for}  \quad}

\author[a]{\large John Joseph M. Carrasco,}
\author[a,b]{\large Matthew Lewandowski,}
\author[a]{\large and Nicolas H. Pavao}

\affiliation[a]{Department of Physics and Astronomy, Northwestern
  University, Evanston, Illinois 60208, USA}
   \affiliation[b]{ Institut fur Theoretische Physik, ETH Zurich,
8093 Zurich, Switzerland}
\emailAdd{carrasco@northwestern.edu, mlewandowski@phys.ethz.ch, 
\\
pavao@u.northwestern.edu}

\begin{document}
\setstackgap{S}{6pt}
\setstackgap{L}{7pt}

\title{\center \fontsize{18.5}{24} \selectfont  Double-copy towards supergravity inflation 
\\
with $\bm{\alpha}$-attractor models}

\abstract{
Key to the simplicity of supergravity $\alpha$-attractor models of inflation are Volkov-Akulov fermions, 
often in the form of nilpotent superfields. Here we explore the possibility of using the double-copy to construct theories of Dirac-Born-Infeld-Volkov-Akulov (DBIVA) coupled to supergravity.   A color-dual bootstrap admits scattering amplitudes involving pions and vectors through five-point tree-level order by order in mass-dimension, but requires the introduction of a $\text{Tr}(F^3)$ operator. Gauge theories with this operator were recently found to require a tower of higher-derivative operators to be compatible with color-kinematics duality.  Adjoint-type double-copy construction at its most conservative seems to require the UV completion of $\text{DBIVA}$ + pure \Poincare supergravity scattering amplitudes to a family of theories involving $\text{DBIVA}$-like particles coupled to Weyl-Einstein supergravity. We also point out an alternative solution to color-dual gauged pions that allows adjoint double-copy without a tower of higher derivative corrections but at the cost of exchange symmetry between scalars. 

}
\makeatletter
\gdef\@fpheader{\,}

\maketitle

\section{Introduction}

Inflation  is a framework involving dynamic spacetime.  Can we begin to understand questions asked of an inflationary universe by considering much simpler questions asked of a gauge theory evolving in flat spacetime?  The existence of color-dual double-copy structure~\cite{BCJ,BCJLoop} for both quantum field theory scattering amplitudes in Einstein-Hilbert gravity, as well as for classical solutions, is incredibly suggestive. For recent reviews of the double-copy literature see refs.~\cite{White:2017mwc,Borsten:2020bgv,BCJreview}.  Here we make some first steps towards a double-copy construction of scattering amplitudes in theories that could support supergravity $\alpha$-attractor models of inflation.

Supergravity $\alpha$-attractor models are among the set of simple yet phenomenologically viable inflationary models consistent with Planck and BICEP (see e.g.~\cite{BICEP:2021xfz, Kallosh:2021mnu} and references therein).  The simplicity of these models relies on  a construction involving nilpotent superfields, which tame the proliferation of unwanted scalar modes during inflation. We have long known that these seemingly esoteric superfields can be understood\footnote{See also ref.~\cite{Bansal:2020krz} for connections made to de Sitter solutions  from a unimodular approach to supergravity~\cite{Nagy:2019ywi}.} as Volkov-Akulov (VA) fermions~\cite{Rocek:1978nb, Lindstrom:1979kq, Casalbuoni:1988xh, Komargodski:2009rz,Antoniadis:2014oya}.  The equivalence of various nonlinear field redefinitions that relate  these representations have been explicitly verified through direct calculation of tree-level scattering amplitudes~\cite{Karlsson:2017bfv}.  So, with eyes on eventually exploring these inflationary models, we can start by introducing double-copy descriptions of Volkov-Akulov fermions with gravity.
 
 To set the stage, we recall a few well-known double-copy constructions.  The first is that dynamical gravity\footnote{To be precise $\mathcal{N}=0$ supergravity is the result of this straightforward double-copy. One can recover Einstein-Hilbert gravity by including additional wrong-sign matter~\cite{Johansson:2014zca, Luna:2017dtq} or by explicitly projecting out unwanted states~\cite{Carrasco:2021bmu}.} can be described by double-copying two Yang-Mills theories, schematically denoted by
 \be
 \mathcal{M}^{\rm gravity} = \mathcal{A}^{\rm YM} \otimes \mathcal{A}^{\rm YM}  \ . 
 \ee
We will explain this construction in detail in \secref{sect:dcReview}, but the main point is that we construct scattering amplitudes in a theory of spin-2 particles by combining simpler building blocks from theories of spin-1 particles.  It is also known that Volkov-Akulov fermions have a double-copy description in terms of nonlinear sigma model (NLSM) scalar pions (denoted here and elsewhere as `$\pi$' for short) and fermions through~
\be
\mathcal{A}^{\rm VA} = \mathcal{A}^{\pi} \otimes \mathcal{A}^{\rm fermions}  \ .
\ee
The above VA fermion amplitudes can be imbedded in a maximally supersymmetric theory by double-copying NLSM pions with the full $\mathcal{N} =4 $ super Yang-Mills (sYM) multiplet, giving a theory of Dirac scalars, VA fermions, and Born-Infeld photons, referred to as Dirac-Born-Infeld-Volkov-Akulov (DBIVA) ~\cite{Cachazo2014xea,He:2016vfi,Cachazo2016njl,Carrasco2016ldy}, 
\be
\mathcal{A}^{\text{DBIVA}} = \mathcal{A}^{\pi} \otimes \mathcal{A}^{\rm sYM}  \ .
\ee

We will build toward a double-copy construction of $\alpha$-attractor models by addressing the possible double-copy structure of scattering amplitudes involving Volkov-Akulov fermions, gravitons, and their superpartners.  We do so by conjecturing the existence of a special color-dual gauge theory of NLSM pions interacting with vectors and attempt to bootstrap it, adding higher-derivative operators as necessary for double-copy consistency~\cite{f3Ladder}.  When double-copied with $\mathcal{N}=4$ super Yang-Mills (sYM) theory, this color-dual vector-pion theory's predictions would presumably generate scattering amplitudes in $\mathcal{N}=4$ Dirac-Born-Infeld-Volkov-Akulov-Supergravity ($\text{DBIVA}+\text{SG}$), schematically
\begin{equation} \label{dcsummary}
\mathcal{A}^{S^2=0} \sim \mathcal{A}^{\text{DBIVA}+\text{SG}} \equiv \mathcal{A}^{\text{YM}+\pi}\otimes\mathcal{A}^{\text{sYM}} \ ,
\end{equation}
where $\mathcal{A}^{S^2=0}$ are S-matrix elements describing nilpotent superfields coupled to dynamic spacetime. In our approach, employing the double-copy construction, the spectrum of the $\mathcal{N}=4$ supergravity theory is realized as a tensor product between the fields of the constituent two gauge theories: a single vector field coupled to NLSM pions and the states belonging to an $\NeqFour$ vector multiplet,
\begin{multline}
(A^+, \lambda^{+}_{ABC}, s_{AB}, \lambda^{-}_{A}, A^-)\otimes (A^+, A^-,\pi)=
(h^{++}, \psi^{+}_{ABC}, A^+_{AB}, \chi^{+}_{A}, \bartaua)\\
\oplus
( \taua, \chi^{-}_{ABC}, A^-_{AB}, \psi^{-}_{A}, h^{--})
\oplus
(A^{\text{BI}}{}^+, \lambda^{\text{VA}}{}^{+}{}_{ABC}, s^{\text{DBI}}{}_{AB}, \lambda^{\text{VA}}{}^{-}{}_{A}, A^{\text{BI}}{}^-)  \ .
\label{dbivaSpectrum}
\end{multline}
All supersymmetry relations in the supergravity theory are inherited from the supersymmetry of the $\mathcal{N}=4$ sYM factors.  Following the conventions of e.g. ref.~\cite{Carrasco:2013ypa}, we use  $(\taua,\bartaua)$ to refer to the  complex  field that labels the external scalar states in 
the supergravity scattering amplitudes. In terms of the two vector fields, these scalar fields are defined by, 
\be
\taua = A^+_{\NeqFour}\otimes A^-_{\text{YM}+\pi} 
\quad , \qquad
\bartaua = A^-_{\NeqFour}\otimes A^+_{\text{YM}+\pi} 
\ .
\label{taua}
\ee
Similarly the double-copy of our conjectured color-dual vector-pion theory with any supersymmetric Yang-Mills theory, including e.g. $\mathcal{N}=1$ sYM, should land on an associated reduced supergravity theory coupled to a likewise reduced DBIVA multiplet. Indeed many reduced symmetry supergravity theories can be considered by orbifolding the supersymmetric gauge theory~\cite{Carrasco:2012ca,Chiodaroli:2013upa} in combination with explicit projecting out of unwanted states~\cite{Johansson:2014zca}.

In this paper we demonstrate how a color-dual bootstrap allows us to build amplitudes in two models of double-copy consistent gauge theories of vectors and pions. In the most conservative approach, we discover that additional higher-derivative operators must be added to the action of a simple covariantized nonlinear sigma model to preserve the color-dual structure beyond four-points. As was found in \cite{f3Ladder}, including these additional operators induces a tower of color-dual contact terms required by the principle of \textit{double-copy consistency}. By collecting terms at each order in the pion decay width, $\Lambda$, we conjecture that the amplitudes can be resummed to a dimensional reduction of $(DF)^2+\text{YM}$ \cite{JohanssonConformal}. We then use the resulting color-dual vector-pion amplitudes in the double-copy construction of Einstein-Weyl-DBIVA theory.  There is a natural question here as to whether the modified Adler's zero, induced by the required higher derivative operators, has allowed us to achieve our aims of actually building nilpotent superfields coupled to supergravity.  This motivates us to identify an alternative  approach to an adjoint color-dual gauged non-linear sigma model with unmodified soft-behavior.  This comes at a cost, however, which is to break manifest exchange symmetry among pions.

The paper is organized as follows. In section \ref{sect:dcReview}, we review the amplitude relations required for double-copy construction and the procedure for constructing double-copy consistent theories using color-dual numerators.  In section \ref{sect:bootstrapF3}, we use an on-shell bootstrap to identify additional operators required for our candidate vector-pion theory to obey color-kinematics duality. In section \ref{sect:bootstrapF3Tower}, we observe that these additional operators induce a tower of higher-derivative corrections, which leads us to conjecture that our desired color-dual vector-pion theory is a dimensional reduction of $(DF)^2+\text{YM}$ theory. In section \ref{sect:novelFeaturesF3}, we identify virtual gluon corrections to the six-point pure-pion amplitudes, required by double-copy consistency in the $(DF)^2+\text{YM}+\pi$ theory, and comment on their implications for the expected Adler's zero of pure Volkov-Akulov observables.  In section \ref{sect:radicalModel} we present a second model for color-dual gauged pions that leaves Adler's zero unmodified at the cost of scalar exchange symmetry.  Our conclusions and discussion of future directions are in section \ref{sect:Conclusions}.

%
%
%

\section{Review of amplitude relations and the double-copy}\label{sect:dcReview}

In this work, we are interested in combining the two theories written schematically on the right-hand side of \eqn{dcsummary}. The key to our procedure for calculating the desired scattering amplitudes of Volkov-Akulov (VA) fermions coupled to gravity is the \textit{double-copy construction} \cite{BCJ}.  Here we provide a brief review of how to implement the double-copy, along with a general overview of on-shell methods.  For more details we encourage the reader to consult recent dedicated review papers~\cite{BCJreview,Adamo:2022dcm}. 

We start by considering a pure $SU(N)$ Yang-Mills theory, whose action is given by 
\be \label{ymaction}
\mathcal{L}^{\rm YM} = - \frac{1}{4} \Tr [F_{\mu \nu} F^{\mu \nu}]  \ , 
\ee
where $F_{\mu \nu} = \partial_\mu A_\nu - \partial_\nu A_\mu - i g [A_\mu , A_\nu ]$ is the field strength, $A_\mu = A_\mu^a T^a$ is the massless gluon field, $T^a$ are the generators of $SU(N)$ (in the adjoint representation), and the trace is over color indices.  The field strength can also be written in terms of the covariant derivative $D_\mu$ as
\be
F_{\mu \nu } = \frac{ i }{g} [ D_\mu , D_\nu] \forr D_\mu = \partial_\mu - igA_\mu \ , 
\ee
where $g$ is the gauge coupling.  The generators satisfy the normalization,
\be 
\label{norm}
[T^a , T^b ] =  f^{abc}T^c  \andd  \Tr[ T^a T^b] = \delta^{ab} \ ,
\ee 
where $f^{abc}$ are the totally antisymmetric structure constants that satisfy the Jacobi identity,
\be \label{jacobiid}
  f^{abe}   f^{ecd} -  f^{dae}   f^{ebc} - f^{dbe} f^{eca} = 0 \ . 
\ee
To compute scattering amplitudes in this theory, one option is to fix a gauge and use the action of \eqn{ymaction} to define Feynman rules. This has the advantage of making locality manifest and attaching factors of $f^{abc}$ to each three-point vertex, but in turn requires ghosts to remove gauge redundancy at loop level, and involves generally laborious calculation.  Off-shell Feynman rules are verbose, whereas physical on-shell quantities can be relatively compact.

Two of the lessons from the on-shell community over the past few decades has been to recycle on-shell information whenever possible through unitarity methods \cite{UnitarityMethod,Fusing}, as well as to minimize the work being done at any given time by considering color-ordered or color-stripped quantities in gauge theories.  We will discuss both, focusing first on color-decomposition.

\subsection{Cubic representations}
Gauge theory scattering amplitudes at tree-level can be expressed in terms of cubic (trivalent) graphs, as in the scaffolding of $\phi^3$ Feynman diagrams,
\be
\mathcal{A}_m = g^{m-2} \sum_{g\in \Gamma_3^m} \frac{ c_g n_g }{ d_g}\,.
\ee
Here $\mathcal{A}_m$ is referred to in the amplitudes community as a full or color-dressed $m$-point or $m$-particle amplitude.  The sum is over $(2m-5)!!$ distinct, connected, cubic tree-graphs with $m$ external legs.  The color-weight $c_g$ of each graph $g$ is calculated by giving each vertex a structure constant $f^{abc}$, and every internal edge a delta function $\delta^{ab}$.  The external edges for these graphs are always labeled, $1,\ldots, m$ and their color indices are in concordance with their labeling, $a_1,\ldots,a_m$.  The propagators or denominator weights, $d_g$ are dressed as normal for massless $\phi^3$ Feynman diagrams, and the kinematic numerator weights $n_g$ are dressed with whatever product of Lorentz invariants (including polarizations for external gluons) are required to correctly reproduce the scattering amplitude.   

The famous Yang-Mills  quartic vertex familiar from Feynman rules is absorbed into this cubic representation by appropriate factors of inverse propagators appearing in the kinematic numerators.  Such assignments are not at all unique, and so there are many distinct representations that are all sufficient.   Indeed one way of generating such a representation is to begin by drawing all cubic Feynman diagrams, dressing them as normal, and then drawing all quartic diagrams, and expanding them out to cubic diagrams consistent with their color-dressings.  This would not be efficient but is certainly a consistent way of generating a cubic representation.

For the three-point amplitude (non-vanishing only for complex momenta), with $k_1+k_2+k_3=0$ and $k_i^2=0$, there is only one distinct graph with no internal propagators,
\be
\mathcal{A}^{\text{YM}}_3 = g \, n_3\, c_3
\ee
with $c_3 = f^{a_1 a_2 a_3}$.  Due to the adjoint structure of $c_3$, the kinematic weight must also be antisymmetric in exchange of particle labels, otherwise the full amplitude would vanish.  Furthermore, there must be one polarization per external leg per term, and the amplitude must be gauge invariant, \textit{i.e.}, $\mathcal{A} |_{\varepsilon_i \to k_i}=0$.  As all Mandelstam invariants $s_{ij}\equiv(k_i+k_j)^2$ vanish at three-points, there are only two possible mass-dimensions one can write down. The lowest order is associated with Yang-Mills, $\text{Tr}(F^2)$:
\be
\label{fSqNumThree}
 n_3 = (\varepsilon_1 \cdot \varepsilon_2)\left( (k_1-k_2)\cdot \varepsilon_3 \right) + \text{cyclic}\,,
\ee
the other is associated with the $\text{Tr}(F^3)$ operator, which it turns out we will have use for in this paper, so we quote its functional form here:
\be
\label{fCubNumThree}
 n^{F^3}_3=  \left( (k_1-k_2)\cdot \varepsilon_3 \right) \left( (k_2-k_3)\cdot \varepsilon_1 \right) \left( (k_3-k_1)\cdot \varepsilon_2 \right)  \,.
\ee
Note that the $n^{F^3}_3$ kinematic weight, when used, must come in with a dimensionful coupling relative to $n_3$.  Again, up to the overall Wilson coefficient of the $\Tr(F^3)$ operator, the kinematic weight of $n^{F^3}_3$ is entirely fixed by mass-dimension and anti-symmetry.  As there are no additional graphs, each of these numerators must be gauge-invariant independently, which indeed they are.

Let us look at a slightly more interesting example, four-points, in Yang-Mills. Here the scattering amplitude is given by,
\be
\label{colorDressedFour}
g^{-2} \, \mathcal{A}^{\text{YM}}_4 = \frac{c_s n_s}{s_{12}} + \frac{c_t n_t}{s_{14}} + \frac{c_u n_u}{s_{13}}\,,
\ee
where $s_{ij}$ is used to mean $(k_i+k_j)^2$.  The color factors are given as follows:
\begin{align}
c_s &= f^{a_1a_2e} f^{e a_3 a_4} \,,\\\,
c_t  &= f^{a_4a_1e} f^{e a_2 a_3}\,, \\
c_u &= f^{a_1a_3e} f^{e a_2 a_4} \,.
\end{align}
The traditional contact term has been absorbed to the cubic dressings by appropriately multiplying $c_s$, $c_t$, or $c_u$  within the contact term  by $s_{12}/s_{12}$, $s_{14}/s_{14}$, and $s_{13}/s_{13}$ respectively.  As the color-weights satisfy the Jacobi identity $c_s=c_t+c_u$ this assignment is by no means unique.  

 One such assignment results in the following functional dressing~\cite{BCJreview} for the kinematic numerators:
\begin{multline}
\label{functionalFourPoint}
n(abcd)  = \Bigl\{ \Big[(\varepsilon_a \cdot \varepsilon_b) k_a^\mu+2 
(\varepsilon_a \cdot k_{b}) \varepsilon_b^\mu-(a \leftrightarrow b)\Big] 
\Big[ (\varepsilon_c \cdot \varepsilon_d) k_{c\,\mu}+2(\varepsilon_c \cdot k_d) 
\varepsilon_{d\,\mu}-(c \leftrightarrow d)\Big]  \\
+ s_{ab} \Bigl[ (\varepsilon_a \cdot \varepsilon_c)(\varepsilon_b \cdot \varepsilon_d)
- (\varepsilon_a \cdot \varepsilon_d)(\varepsilon_b \cdot \varepsilon_c)\Bigr] \Bigr\}\,,
\end{multline}
with
\begin{align}
 n_s &= n(1234)\, , \\
 n_t &= n(4123)\, ,\\
 n_u &= n(1324)\, .
\end{align}
In particular it is worth noticing that the coefficient of $s_{ab}$ in \eqn{functionalFourPoint} is exactly the piece of the contact term absorbed by each cubic graph.

\subsection{Color decomposition}
At four-points, as well as higher multiplicity, local dressings of individual graphs are no longer individually gauge invariant. Why? There can be cancellations between graphs depending upon the redundancy between color factors.   Generically the color-factors satisfy the Jacobi identity,
\be
  f^{abe}f^{ecd}=f^{dae}f^{ebc} + f^{cae}f^{edb}
\ee
with an implied sum over the repeated indices. Because of this redundancy in color-factors, any representation that makes use of all $(2m-5)!!$ cubic graphs and their individual color-dressings are not in a minimal color basis.   One can always use the Jacobi identities to re-express the color dressings in a minimal color basis of $(m-2)!$ color weights~\cite{DixonMaltoni}, and the kinematic coefficients of those color weights will be gauge invariant.

Let us look at the four-point example above to see a concrete realization.  One can rewrite \eqn{colorDressedFour} using $c_t=c_s-c_u$ to achieve,
\begin{equation}
\label{cdFourPointAmps}
 g^{-2}  \, \mathcal{A}^{\text{YM}}_4 = c_s \left( A^{\text{YM}}_{4}(1234)  \equiv \frac{ n_s}{s_{12}}  + \frac{ n_t}{s_{14}} \right) + c_u  \left( A^{\text YM}_{4} \equiv \frac{n_u}{s_{13}}- \frac{ n_t}{s_{14}}\right)\,.
\end{equation}
The color-ordered  or partial amplitudes $A^{\text{YM}}_{4, \text{tree}}(1234)$ and $A^{\text{YM}}_{4, \text{tree}}(1324)$ are gauge invariant and only depend on an exponential number of cubic graphs as multiplicity increases~\cite{BCJreview}.  They are related functionally and satisfy various field theory relations between them.   

\begin{figure}[t]
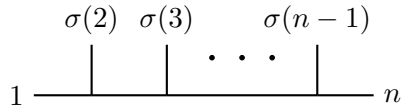

\begin{center}
\halfLadder{}
\end{center}
   \caption{ \label{fig:halfLadderDiagram} The basis graphs for any tree-level kinematics that satisfy Jacobi-like relations can be expressed in terms of {\em half-ladder diagrams} where two legs, say $1$ and $n$, are held fixed, and all other leg labels $\sigma(i)$ are permuted. See, e.g., ref.~\cite{DixonMaltoni} for the relevant argument applied to color-factors.}

\end{figure}
Generically, we can write $m$-multiplicity gauge theory scattering amplitudes in the following form:
\begin{align}
 \label{ymtree2}
\mathcal{A}^{\rm YM}_{n, \text{ tree}}  &=  g^{n-2} \sum_{g_i \in \Gamma_3^{(n)}} \frac{c_{g_i} \, n_{g_i}}{d_{g_i}} \\
 \label{ymtree3}
&= g^{n-2} \sum_{\sigma \in S_{n-2}}   c(1,\sigma(2),\dots ,\sigma(n-1),n)   A_{n}^{\rm YM} (1,\sigma(2), \dots , \sigma(n-1) , n    ) \ .
\end{align}
where the $c(1,\sigma(2),\dots ,\sigma(n-1),n)$ corresponded precisely to a product of color factors dressing a half-ladder or multiperipheral graph,
\begin{equation}\label{eq:colorDef}
c(1,\sigma(2),\dots ,\sigma(n-1),n) \equiv  f^{a_1 a_{\sigma(2)} b_1 } f^{b_1 a_{\sigma(3) } b_2 } \cdots  f^{b_{n-3} a_{\sigma ( n-1)} a_n } \ .
\end{equation}
Given this color structure, we call \eqn{ymtree3} the half-ladder representation, as it restructures the color-dressed amplitude in terms of a sum over color factors associated with ``half-ladder" cubic graphs, see e.g.~\Fig{fig:halfLadderDiagram}. In the case of \eqn{eq:colorDef}, root legs $1$ and $n$ are held fixed, and all other legs take on permutations of labels $\sigma$. The remaining color dressings are then related to this choice of $c(1,\sigma(2),\dots ,\sigma(n-1),n)$ by iterated application of Jacobi identities on the internal edges, or antisymmetry of individual vertices \cite{DixonMaltoni},
\begin{equation}\label{eq:jacobiDiagram}
\tchannel =\schannel - \uchannel\,,
\end{equation}
where $a_i$ are labels of internal or external edges. The half-ladder basis of \eqn{ymtree3} is expressed in terms of a minimal color basis, meaning each kinematic partial amplitude contributes to the overall amplitude weighted by a completely distinct direction in color space.
Before proceeding, it is worth taking a moment to appreciate the utility of reframing gauge invariant amplitudes in this way. While the local behavior manifest in the Feynman rules generated \eqn{ymaction} has been obscured (at least for now), the local gauge symmetry and Bose symmetry is now absolutely manifest.

One can choose alternative color-bases depending on which external legs are held fixed as the half-ladder root legs. Their coefficients are also gauge invariant and indeed the partial amplitudes of one basis can be linearly expressed in terms of partial amplitudes of other bases. This is due to a set of linear relations known as the Kleiss-Kuijf relations \cite{Kleiss:1988ne}, which follow simply from the fact that kinematic dressings of graphs obey antisymmetry around vertices, correspondent with the color-weights, to preserve Bose symmetry. The partial amplitudes also inherit reflection and cyclic identities,
\be
A_{n,\text{tree}}^{\rm YM} (1 , 2 , \dots , n)= A_{n,\text{tree}}^{\rm YM} ( 2 , \dots , n , 1) \ , 
\ee
and 
\be
 A_{n,\text{tree}}^{\rm YM} ( 1 , 2 , \dots , n ) = (-1)^n A_{n,\text{tree}}^{\rm YM} (n , \dots , 2 , 1) \ ,
\ee 
and also obey what is called the photon decoupling identity,
\be
\sum_{\sigma \in \text{cyclic}} A_{n,\text{tree}}^{\rm YM} ({ \sigma(1), \sigma( 2) ,\dots , \sigma(n-1)},n) = 0 \ . 
\ee

As we will now show, the partial amplitudes respect an additional set of relations that underly the local properties of the kinematic numerators hidden in each $ A_{n,\text{tree}}^{\rm YM}$.

%
%

\subsection{Color-dual numerators}

Let us spend a second revisiting the generalized gauge choices that allow us to make different assignments of contact terms to cubic graphs.  As we mentioned earlier, when a higher-point contact interaction, $X_g$, weighted by a color factor $c_g$, is produced by the Feynman rules, it can be made compatible with the form of \eqn{ymtree2} by multiplying by an overall factor of unity, taking $c_g X_g \rightarrow \frac{c_g }{d_g}(d_g X_g)$. In which case, $n_g \rightarrow n_g + d_g X_g$.  Take for example any three graphs $\{g_1 , g_2 , g_3\}$ in $\Gamma_3^{(m)}$ whose color factors are related by the Jacobi identity
\be
c_{g_1}  = c_{g_2} + c_{g_3} \ . 
\ee
Then the amplitude is invariant under shifting the numerator factors by the following \emph{generalized gauge transformation}
\be
n_{g_1} \rightarrow n_{g_1} + d_{g_1} \Delta_{\{g_1 , g_2 , g_3\}}  \ , \quad n_{g_2} \rightarrow n_{g_2} - d_{g_2} \Delta_{\{g_1 , g_2 , g_3\}}  \ , \quad n_{g_3} \rightarrow n_{g_3} - d_{g_3} \Delta_{\{g_1 , g_2 , g_3\}}  \ , 
\ee
leading to an overall shift in the amplitude of,
\be
\mathcal{A}^{\rm YM}_{n, \text{ tree}} \rightarrow \mathcal{A}^{\rm YM}_{n, \text{ tree}} + (c_{g_1}-c_{g_2}-c_{g_3})\Delta_{\{g_1 , g_2 , g_3\}} = \mathcal{A}^{\rm YM}_{n, \text{ tree}} \,
\ee
for any function $\Delta_{\{g_1 , g_2 , g_3\}}$ of the kinematic invariants. This is equivalent to the process of absorbing contact terms into the trivalent graph representation of the full amplitude. 

In the case of Yang-Mills amplitudes, for all graphs whose color factors are related by the Jacobi identity, these $\Delta$ can be chosen such that the $n_g$ satisfy the same relationships as the color factors $c_g$, i.e.
\begin{align}
\begin{split} \label{colordualform}
 c_{g_i} + c_{\bar{g}_i}=0 \quad & \Rightarrow \quad  n_{g_i} + n_{\bar{g}_i}=0  \ , \\
 c_{g_i} - c_{g_j} - c_{g_k} = 0 \quad & \Rightarrow \quad n_{g_i} - n_{g_j} - n_{g_k} = 0 \ , 
\end{split}
\end{align}
where $\bar{g}$ is a cubic graph with an odd number of vertex flips relative to $g$. In this gauge, the interplay between the algebraic relations of the local numerators and the color factors now make Bose symmetry and gauge invariance manifest. Given the parallel structure between the color and kinematic factors, we say that Yang-Mills theory obeys the \textit{duality between color and kinematics}, or simply, that it is \textit{color dual}.

A wide variety of theories have been shown in the literature to be color dual, as recently reviewed in refs.~\cite{BCJreview, Carrasco:2020ywq,Adamo:2022dcm}. Such theories permit this special representation of the numerators if and only if the partial amplitudes $A_{n,\text{tree}}^{\rm YM}$ satisfy a further set of relations, identified by Bern, Johansson and one of the authors \cite{BCJ}, 
\be\label{eq:BCJrel}
\sum_{i = 2}^{n-1} k_1 \cdot \left(  k_2 + \dots + k_i \right)  A_{n,\text{tree}}^{\rm YM} (2 , \dots , i , 1 , i+ 1 , \dots , n) = 0 \ ,
\ee
sometimes called the fundamental BCJ relations~\cite{Feng:2010my}. These additional kinematic identities further reduce the basis of partial amplitudes from $(n-2)!$ to $(n-3)!$ elements. This leads to an associated reduction in complexity of constructing both tree-level amplitudes and multi-loop integrands via cut-construction.
%
%

\subsection{The double-copy}
A reduction in computational complexity is not the only advantage of performing calculations in color-dual theories. Consider the replacement of color-weight with color-dual vector kinematic weight, $c_g \rightarrow n_g$,  yielding,
\be \label{dcexample}
\mathcal{M}_{n,\text{tree}} = \sum_{g_i \in \Gamma_3^{(n)}}   \frac{ n_{g_i} n_{g_i}  }{d_{g_i}}  \ . 
\ee
The resulting theory is local by construction and remains Bose symmetric due to the algebraic properties of the numerators and the explicit sum over all permutations of graph labels. But now the amplitudes are colorless with states determined by the tensor product $\varepsilon_i^\mu \otimes \varepsilon_i^\nu$. Furthermore, the new amplitude is invariant under \emph{linear diffeomorphisms}, the ``square" of gauge invariance. In other words, two factors of Yang-Mills theory numerators yields a theory of axion-dilaton-gravity. We call this procedure the \textit{double-copy construction}.

More generally, we can be agnostic to the origin of the numerators, and construct a general class of theories of the form
\be \label{DCCex}
\mathcal{M}_{n,\text{tree}} = \sum_{g_i \in \Gamma_3^{(n)}}   \frac{ a_{g_i} b_{g_i}  }{d_{g_i}}  \ . 
\ee
where the numerator weights $a_{g_i}$ and $b_{g_i}$ are local functions of {color} and kinematics that obey the same algebraic relations of \eqn{colordualform}, subject to a suitable gauge choice. As long as both the $a$-stripped ordered amplitudes and $b$-stripped ordered amplitudes are consistent across multiplicities via factorization, we call such theories \textit{double-copy consistent}. By construction they are local and invariant under generalized gauge transformations acting on each set of numerators.  

It is worth also briefly mentioning how this process continues to loop level.  In Yang-Mills, at $L$ loop level, the $n$-point amplitude in $D$ dimensions is given by,
\be
\mathcal{A}^{\rm YM}_{n,L\text{-loop}} = i^{L-1} g^{n-2 + 2L} \prod_{l=1}^L  \int \frac{d^D \ell_l }{(2 \pi)^{D}} \sum_i \frac{1}{S_i} \frac{c_i n_i}{d_i }
\ee
where $\ell_l$ are the loop momenta, the sum is over all cubic diagrams that can be built from Feynman rules with distinct color structure $c_i$, the $d_i$ are products of propagators for each internal line. The $n_i$ depend on contractions of the various kinematic quantities (including loop momenta), and $S_i$ are symmetry factors.  Then, the corresponding gravitational theory has loop amplitudes \cite{BCJ,BCJLoop} of the form,
\be
\mathcal{M}_{n,L\text{-loop}} = \mathcal{A}^{\rm YM}_{n,L\text{-loop}} \Big|_{c_i \rightarrow \tilde n_i , g \rightarrow \kappa / 2} = i^{L-1} \left( \frac{\kappa}{2} \right)^{n - 2 + 2 L }   \prod_{l=1}^L  \int \frac{d^D \ell_l }{(2 \pi)^{LD}} \sum_i \frac{1}{S_i} \frac{\tilde n_i n_i}{d_i } \ . 
\ee

As noted previously, the above considerations naturally extend to theories other than Yang-Mills.  The key to the double-copy in all of these cases is finding color-dual kinematic numerators $n_g$, i.e. that satisfy the Jacobi relations \eqn{colordualform}.  The goal in this paper is to double-copy sYM with a color-dual theory of vectors coupled to pions, which we will build up using an on-shell bootstrap. Once we have done so, we can replace adjoint color factors with color-dual sYM numerators, thereby constructing the desired observables in a putative theory of VA fermions coupled to supergravity.  

\subsection{Unitarity methods}\label{sec:unitarityMethods}
We presented gauge-invariant building blocks of gauge theories in terms of easier to compute, purely kinematic partial amplitudes, and we saw how it was possible to have algebraic constraints on the building blocks of such partial amplitudes by imposing color-kinematics duality on the numerator factors.   How do we know that these partial amplitudes are consistent with the theory we mean to be talking about?  Generally they must satisfy what are called unitarity cuts.  At tree-level, this is simply factorization to lower-multiplicity on-shell amplitudes, summed over states in the cut.   The requirement of such consistency allows for verification, and verification is sufficient for construction, as we will shortly discuss.  Furthermore, by considering ordered amplitudes that contribute to \textit{ordered-cuts}, far fewer graphs will be needed for any given cut.

To proceed, we note that the sum of scalar states is trivial, and that the sum over massless vector states is given by the following completeness relation in terms of formal polarizations on a cut leg with momenta $k$:
\begin{align}
\label{gluonCompleteness}
\sum_{s \in \text{states}} \varepsilon_s^\mu (k) \varepsilon_{\bar{s}}^\nu(-k) = \eta^{\mu \nu} - \frac{k^{\mu} q^{\nu} + q^{\mu} k^{\nu}}{k\cdot q}\,,
\end{align}
 where $q$ is an arbitrary null reference momentum that must cancel out in any physical quantities, such as ordered amplitudes. 
 
As an example, consider the $s_{12}$-channel cut of the partial amplitude $A^{\text{YM}}(1234)=n_s/s_{12} + n_t/s_{14}$.  We are looking to verify,
\be
 \lim_{s_{12}\to0} s_{12} A(1234) =  \sum_{s\in \text{states}} A(1,2,l_s)A(-l_{\bar{s}},3,4)
 \ee
where $A(abc)=n_3(abc)$ is given in \eqn{fSqNumThree}. From the LHS of our cut equation we have:
\begin{align}
 \lim_{s_{12}\to0} s_{12} A(1234) &= \left. n_s  \right |_{s_{12}\to0}  \nn \\
&= \left. n(1234)  \right |_{s_{12}\to0} \nn \\
&= \left. A(1,2,l)A(-l,3,4) \right|_{\varepsilon(l)^{\mu} \varepsilon(-l)^\nu \to \eta^{\mu \nu} } \,.
\end{align} 
In the last line we have exploited the property that --- on the cut --- the contact term drops out of \eqn{functionalFourPoint}, and the remaining inner product in the numerator can be written as a product between two on-shell three-point amplitudes if the cut-polarizations are interpreted as a spacetime metric.  This is manifestly equal to the RHS of our cut equation -- since in \eqn{gluonCompleteness}, only the metric contributes in our sum over states, as $k_i\cdot l=0$ for all external legs $i$ due to three-point on-shell kinematics \cite{Kosmopoulos:2020pcd}.

%
%
\subsection{Unitarity based bootstrap}
\newcommand\cS{{(D\phi)^2}}

One can invert verification to construction in the form of unitarity-based bootstrap methods \cite{UnitarityMethod,Fusing}.  This is a longstanding approach applied to both tree amplitudes and loop-level integrands.  We will focus here on the idea of requiring the duality between color and kinematics in conjunction with unitarity based factorization.

Let us consider as an example a bootstrap through four-point for the amplitudes associated with the covariantized free scalar: $\mathcal{L}= (D\phi)^2 + \text{Tr}(F^2)$.  We have already fixed the purely gluonic sector, so we are left to consider amplitudes involving scalars. Consider the three-point amplitude between adjoint scalars and a massless vector. The color weight will be of the form $f^{abc}$ and so the kinematic weight is incredibly restricted.  As all dot products between  on-shell external momenta vanish for three-point kinematics, the only three-point amplitude one can write down is:
\be
\label{covScalarThree}
\mathcal{A}^{\cS}_3(1,2,3_A) = g  f^{abc} (k_1-k_2)\cdot \varepsilon_3\,.
\ee
Strictly speaking we could have given a new coupling to the three-point amplitude at this stage, but consistency at four-point will force the same coupling as the three-point gluon.

At four-points, the first and simplest amplitude to consider is one with all-scalar external states. The full four-point amplitude will be of the form,
\be
\label{cdCovScalarFour}
\mathcal{A}^{\cS}_4(1,2,3,4) = g^2 \left( \frac{n_s c_s}{s_{12}} + \frac{n_t c_t}{s_{14}} +\frac{n_u c_u}{s_{13}}  \right) \,.
\ee
The kinematic weight will be linear in Mandelstam invariants if it is to factorize to \eqn{covScalarThree}.  Requiring the duality between color and kinematics lands us uniquely on the kinematic weight up to a constant of proprtionality, $\xi$, 
\be
n^{(D\phi)^2}(ijkl)= \xi \, ( s_{ik} - s_{jk} )\,.
\ee
Here $n_s=n^{(D\phi)^2}(1234)$ and others following standard relabeling as for the gluonic numerators previously constructed in \eqn{functionalFourPoint}. Factorization on the $s_{12}$-channel requires
\begin{align}
 n^{(D\phi)^2}(1234)|_{s_{12}\to0} &= \left. \xi ( s_{13}-s_{23} ) \right|_{s_{12}\to0}\\
  &= 2 \, \xi  \,s_{13} \\
  &= \sum_s A(1,2,l_{A,s}) A(-l_{A,\bar{s}},3,4)\\
  &= \left.(k_1-k_2)\cdot(k_3-k_4)\right|_{s_{12}\to0}\\
  &= 4 \, s_{13} \,
\end{align}
fixing the constant of proportionality $\xi =2$.  The only remaining amplitude to consider through four-points is the two-gluon two-scalar amplitude.  In this case we actually have two distinct graph topologies -- one where each of the external gluons interacts with a separate vertex connected by a scalar propagator, $n_{\text{I}}$, and one where the two gluons interact at the same vertex, $n_{\text{II}}$.  When legs labeled 1 and 4 are vectors, we have,
\begin{align} 
\mathcal{A}^{\cS}(1_A,2,3,4_A) &=  g^2 \left(  \frac{ c_s n_{\text{I}}(1_A,2,3,4_A) }{s_{12}} + \frac{c_t n_{\text{II}}(1_A,4_A,3,2) }{s_{14}} + \frac{ c_u n_{\text{I}}(1_A,3,2,4_A) }{s_{23}}  \right)\\
 &= g^2 \left(  c_s \left[ A(1234) \equiv \frac{ n_{\text{I}}(1_A,2,3,4_A) }{s_{12}} + \frac{ n_{\text{II}}(1_A,4_A,3,2) }{s_{14}} \right]  \right. \nn\\
&~~~~~  \left. + \,\,c_u \left[ A(1324) \equiv \frac{ n_{\text{I}}(1_A,3,2,4_A) }{s_{12}} + \frac{ n_{\text{II}}(1_A,4_A,2,3) }{s_{14}} \right]   \right)\,.
\end{align}
The duality between color and kinematics requires that:
 \be
 n_{\text{II}}(1_A,4_A,3,2) = n_{\text{I}}(1_A,2,3,4_A) -  n_{\text{I}}(1_A,3,2,4_A) \,,
\ee
which is manifestly antisymmetric under exchange $j\leftrightarrow k$. Thus, we need only give an ansatz to $n_{\text{I}}$.  At this mass-dimension, a spanning set of independent Lorentz invariants is simply:
\be 
\left\{ \varepsilon_1 \cdot \varepsilon_4 , \varepsilon_{1}\cdot k_2, \varepsilon_{1}\cdot k_4,
   \varepsilon_{4}\cdot k_1, \varepsilon_{4}\cdot k_2, s_{13}, s_{14} \right\} \,
\ee 
and our initial ansatz is thus:
\begin{multline}
n_{\text{I}}(1_A,2,3,4_A) =  \varepsilon_1 \cdot \varepsilon_4 ( \xi_1 s_{13} + \xi_2 s_{14} ) + (\varepsilon_{1}\cdot k_2) \left(  \xi_3 ( \varepsilon_{4}\cdot k_1) + \xi_4    ( \varepsilon_{4}\cdot k_2) \right )\\
+ (\varepsilon_{1}\cdot k_4) \left(  \xi_5 ( \varepsilon_{4}\cdot k_1) + \xi_6    ( \varepsilon_{4}\cdot k_2) \right )
\end{multline}
Fixing on the $s$-channel ordered cut of $A(1234)$,
\be
 n_{\text{I}}(1_A,2,3,4_A)|_{s_{12}\to0} = A(1_A,2,l) A(-l,3,4_A) = -4 ( \varepsilon_1 \cdot k_2) (\varepsilon_4 \cdot k_2)\,,
\ee
yields $\xi_1=\xi_2$, $\xi_4=4$, and $\xi_3=\xi_5=\xi_6=0$. One can fix the remaining parameters by either considering the $t$-channel cut,
\be
n_{\text{II}}(1_A,4_A,3,2)|_{s_{14}\to0} = \sum_s A(1_A,4_A,l_{A,s})A(-l_{A,\bar{s}},3,4)\,,
\ee
 or demanding gauge invariance of $A(1234)$. Either of these constraints fixes $\xi_2=1$. This yields ultimately,
 \be
 n_{\text{I}}(1_A,2,3,4_A) = - 4 (k_2 \cdot \varepsilon_1) ( k_3 \cdot \varepsilon_4) - s_{12} (\varepsilon_{1}\cdot\varepsilon_{4})\,,
\ee
and we are done through four-points. Note that the presence of $s_{12}$ means that our bootstrap yields the contact-term required for the gauge-invariance of $(D\phi)^2$. We will employ a similar strategy to attempt to couple pions to vectors in a color-dual fashion.

 \subsection{Vector tower coming from $\Tr(F^3)$}
 \label{sect:vectorTower}
  We will see that the $\Tr(F^3)$ operator must be included in our first candidate color-dual vector-pion theory, and this has consequences.  Here, we briefly review the relevant results of \cite{f3Ladder} which described the color-dual bootstrap for a pure-vector Yang-Mills theory deformed by $\Tr(F^3)$, which ultimately must be a part of the vector-scalar theory that we consider in this work.  It was shown in \cite{f3Ladder} that an action of the form\footnote{{To compare the notation of \cite{f3Ladder} to the current paper, we note that \cite{f3Ladder} had $ g = 1$.  Restoring powers of $g$, the coupling of $\Tr ( F^3)$ should be $g \alpha'$ as in \eqn{f3ladderaction}, which  allows us to match the two notations with the identification $\alpha' = - \Lambda / g^2$.}}  
\be \label{f3ladderaction}
\mathcal{L}_{\text{YM} + F^3} = - \frac{1}{4} \Tr (F^2) + \frac{g \alpha '}{3}  \Tr (F^3)  \ , 
\ee
while obviously gauge invariant, is not color dual to all tree-level multiplicity at all orders in $\alpha '$.  Instead, it appears that in order to make the theory double-copy consistent, one must add an infinite tower of higher-derivative operators with increasing powers of $\alpha'$.  This is required even at low multiplicity.  
\begin{figure}[h]
    \centering
    \begin{equation*}
\begin{split}
A^{(n)}{(12345)}\Big|_{s_{45}-\text{cut}}=\fivegraphGluePropOp{1}{2}{4}{5}{\alpha'^{n-1}}{\alpha'^1}{3}{hred}{hgrey1}+ \fivegraphGluePropOp{1}{2}{4}{5}{\alpha'\,^n}{\alpha'^0}{3}{hblue}{hgrey2}  {\mathcal{O}(\alpha'^n)}
\end{split}
\end{equation*}
   \begin{equation*}
\begin{split}
A^{(n+1)}{(12345)}\Big|_{s_{45}-\text{cut}}=\fivegraphGluePropOp{1}{2}{4}{5}{\alpha'\,^{n}}{\alpha'^1}{3}{hred}{hgrey2}+ \fivegraphGluePropOp{1}{2}{4}{5}{\alpha'^{n+1}}{\alpha'^0}{3}{hblue}{hgrey3} \!\!\!\! {\mathcal{O}(\alpha'^{n+1})}
\end{split}
\end{equation*}
    \caption{ \label{fig:vectorTower}Inductive ladder of higher-derivative contacts for the pure-vector theory generated by cuts of the form $A^{(k)}_5|_{\text{cut}}\sim(A_4^{(k)} A_3^\text{YM}+A_4^{(k-1)}A_3^{F^3})$ and considering double-copy consistency, where $A_n^{(k)}$ denotes $n$-point amplitudes appearing at mass-dimension $\alpha'^k$.}
\end{figure}

Consider the four-point amplitude.  The action in \eqn{f3ladderaction} gives a four-point amplitude at $\mathcal{O}(\alpha'{}^2)$ by combining two $ \alpha' \Tr (F^3)$ vertices.  This amplitude is not color dual by itself, but can be made so by adding a term proportional to $\alpha'{}^2 \Tr (F^4)$ to the theory \cite{Broedel2012rc}.  The new $\alpha'{}^2 \Tr (F^4)$ term then contributes to the five-point amplitude.  In particular, there is an $\mathcal{O}(\alpha'{}^3)$ contribution at five points coming from combining $\alpha'{}^2 \Tr (F^4)$ with $\alpha '  \Tr (F^3)$.  This contribution was also shown to not be color dual by itself.  To make it color dual, one has to add a new operator at four points, schematically $\alpha'{}^3  \Tr ( \partial^2 F^4)$, which combines with the $\mathcal{O}(\alpha'{}^0)$ three-point vertex in \eqn{f3ladderaction} to give an $\mathcal{O}(\alpha'{}^3)$ contribution at five points and make the amplitude color dual at that order. However, the new $\alpha'{}^3  \Tr ( \partial^2 F^4)$ can now combine with $ \alpha' \Tr (F^3)$ to give an $\mathcal{O}(\alpha'{}^4)$ contribution at five points, which again is not color dual, but can be fixed by a term schematically like $\alpha'{}^4  \Tr ( \partial^4 F^4)$.  The essence of this cut-consistency is outlined in \Fig{fig:vectorTower}. Thus, by just considering the four- and five-point amplitudes, this tower was conjectured to continue indefinitely, requiring an infinite number of higher-derivative operators to be color dual and satisfy the correct factorization.

%
%
%

\section{Bootstrapping to a color-dual vector-pion theory}\label{sect:bootstrapF3}
One might expect that a color-dual theory involving vectors and pions would be an inevitability. After all, both pure NLSM amplitudes~\cite{Chen2013fya,Cachazo2014xea} and pure gluonic amplitudes are double-copy consistent~\cite{KiermaierTalk,BjerrumMomKernel}
to all multiplicity.  While one can be inspired by color-dual all-vector amplitudes and all-pion amplitudes, the existence of a double-copy consistent theory admitting mixed pion and vector amplitudes is an open question that we will now begin to address.

%

\subsection{Gauged nonlinear sigma models}
A natural first step would be to  couple Yang-Mills to the NLSM via simple covariantization, by promoting $\partial_\mu \rightarrow D_\mu$.    
Consider then the covariantized NLSM Lagrangian\footnote{Starting from the form of the pion action given in e.g.~\cite{Cachazo2014xea}.} for $SU(N)$ adjoint scalars $\pi = \pi^a T^a$ coupled to gluons $A_\mu = A_\mu^a T^a$, 
\begin{equation} 
\label{gaugeNLSMaction}
\mathcal{L}^{\cP} = -\frac{1}{4} \Tr  ( F^2 ) +
\frac{1}{2} \Tr\left[ \left( 1- \Lambda \pi^2 \right)^{-1} D_\mu \pi  \left(  1-\Lambda \pi^2 \right)^{-1} D^\mu \pi  \right] \,,
\end{equation}
where $\Lambda$ is related to the pion decay constant $f_\pi$ by $\Lambda \sim 1 / f_\pi^2$, and the covariant derivative is given as
\begin{equation}
    D_\mu \pi = (\partial_\mu \pi^a -igf^{abc}A_\mu^b \pi^c)T^a \, .
\end{equation}
Such theories, also with additional operators, have a venerable history in the literature, see  refs.~\cite{Brezin:1980ms,Gates:1984nk} for some earlier examples, and are referred to generally as gauged nonlinear sigma models (e.g.~\cite{Hull:1989jk}).  

Our approach is as follows.  We take \eqn{gaugeNLSMaction} as our starting point for a theory of vectors and scalars.  We  then look at amplitudes of increasing multiplicity and check whether they are color dual.  Finding that at five-points they are not, we will modify \eqn{gaugeNLSMaction} using the color-dual bootstrap, essentially adding new operators to \eqn{gaugeNLSMaction}, to ensure that the resulting theory is double-copy consistent.  In the end, we find towers of higher-derivative operators that must be added to \eqn{gaugeNLSMaction} in order for the amplitudes to be color dual-and satisfy consistent factorization to lower-point amplitudes.  

All tree-level amplitudes in this theory through four-points are already known to be color dual. 
The all-vector amplitudes belong to the predictions of Yang-Mills theory. 
The  $\mathcal{O}(\Lambda)$ contribution to the all-scalar four-point amplitude belongs to the predictions of NLSM. The remaining contributions through four point belong to the scattering amplitudes of the covariantized free-scalar theory, sometimes referred to 
as the simple-scalar, i.e. $\Lambda \to 0$ in \eqn{gaugeNLSMaction}.  
The first opportunity for an  amplitude original to the theory \eqn{gaugeNLSMaction} is at five points, which we discuss in the following section.

%
%
%
\subsection{$\Tr(F^3)$ from double-copy consistency}\label{sect:trF3Intro}
To begin, we look at the $\mathcal{O}(g \Lambda)$ behavior of $\mathcal{A}^{\cP}(\pi\pi\pi\pi A)$, with a factorization channel involving the characteristic four-pion amplitude as well as the three-point scalar-vector interaction.   Through explicit calculation from \eqn{gaugeNLSMaction} we find that the tree-level five-point amplitude $\mathcal{A}^{\cP}(\pi\pi\pi\pi A)$, fails to satisfy the consistency relations of ref.~\cite{BCJ} required to participate in standard double-copy construction. 
Recall, however, that we only care about the theory of \eqn{gaugeNLSMaction} as a starting point.  
There are any number of gauge-invariant operators one might introduce, and indeed may be required, for the full theory to be double-copy consistent, as described in \cite{f3Ladder}.

There are two approaches to resolving this issue that we follow.  A potentially more radical solution is to admit the possibility that pions may belong to possibly distinct fields -- breaking exchange symmetry between scalars.  We leave the discussion of such a model to \sect{sect:radicalModel}. The more conservative approach, which we consider now, is to simply admit higher derivative corrections to the fields we have already present in our theory. To identify the requisite operators, we follow the color-dual bootstrap approach of starting with an ansatz for the amplitude of the appropriate kinematic mass-dimension and little-group scaling. We then establish whether or not it is possible to fix parameters so that our desired scattering amplitude is color dual and satisfies the generalized unitarity cuts we value. Previous examples in the literature of a similar approach include bootstrapping NLSM pion amplitudes in ref.~\cite{BCJreview} and massive scalar QCD in ref.~\cite{Carrasco:2020ywq}. In these cases, at a given kinematic mass-dimension, satisfactying color-kinematics duality and factorization alone entirely constrain the amplitudes of the theory.

In the current problem of identifying the existence and nature of a not-completely specified theory, we recognize that we might not have included at lower multiplicity all the interactions required to admit  both color-kinematics duality as well as healthy factorization at five-points.  So as far as factorization is concerned,  we will at first only insist that our five-point amplitude of the correct mass-dimension must be color dual, and satisfy the factorization to four-pion scattering  
\begin{equation}
\label{fourPionCut}
 \mathcal{A}_5(\pi\pi\pi\pi A)|_{(k_4+k_5)\text{-cut}} = \mathcal{A}_4(\pi\pi\pi\pi) \times \mathcal{A}_3(\pi\pi A)\,.
\end{equation}
We will learn what was missing from the simple covariantization of \eqn{gaugeNLSMaction} by probing the analytic structure of this new color-dual amplitude.

Five-point amplitudes with four external scalars can be expressed in terms of kinematic and color dressings of the fifteen distinctly labeled cubic (trivalent) graphs, $\Gamma^{(5)}_3$,
\begin{equation}
\label{fivePoint}
\mathcal{A}_5(1_\pi,2_\pi,3_\pi,4_\pi,5_A)=\sum_{g\in \Gamma^{(5)}_3}\frac{ c_g n_g}{d_g}\,.
\end{equation}
The color-weights $c_g$ are given by dressing each vertex with adjoint structure constants, $f^{abc}$, the propagators $d_g$ are dressed as normal for massless particles, and any remaining kinematics and coupling constants are absorbed into the kinematic numerators, $n_g$.   
  
 The fifteen graphs contributing to the five-particle amplitude can be described as relabelings of only two graph topologies when one encodes external particle type in the structure of the graph.  In our case the four equivalent particles are the pions, and the lone distinct fifth particle is the vector. The basis topology graph, with external gluon on a terminal vertex, we will refer to as $g_1$,
\begin{equation}
g_1(1_\pi,2_\pi,3_\pi,4_\pi, 5_A)=\fivegraphBasis{1}{2}{3}{4}{5}\,.
\end{equation} 
The other topology, $g_2$,  has the external vector on the central vertex,
\begin{equation}
g_2(1_\pi,2_\pi,3_\pi,4_\pi,5_A)=\fivegraphChild{3}{4}{5}{1}{2}\,.
\end{equation}
  Note in neither topology do we bother assigning a particle type to the internal legs (represented as dashed edges here). All propagating particles are massless and it does not affect dressing the graph with kinematic weights, color-weights, or its propagators.   Contact terms are included by allowing for terms proportional to inverse propagators in the kinematic numerators of the cubic graphs.
  
   A color-dual kinematic Jacobi relation gives the kinematic weight of the second topology, $n(g_2)$, in terms of two labelings of the first, $n(g_1)$,
\begin{equation}
 n(g_2(1_\pi,2_\pi,3_\pi,4_\pi,5_A))= n(g_1(1_\pi,2_\pi,3_\pi,4_\pi,5_A)) - (3 \leftrightarrow 4) \,.
\end{equation} 
All fifteen graphs in \eqn{fivePoint} are thus given by the kinematic weight of various relabelings of the basis graph  $g_1$, or linear combinations thereof.    Thus, we need only give an ansatz of the correct mass-dimension and little-group scaling to $n(g_1)$.  Little group scaling demands a polarization for the gluon, $\varepsilon_5$, in each term.  For the $\mathcal{O}(g \Lambda)$ contribution to $\mathcal{A}_5 ( \pi \pi \pi \pi A)$, we require $n(g_1)$ to be third-order in Lorentz-invariants, $(k_i\cdot k_j)(k_l \cdot k_m)(k_n \cdot \varepsilon_5)$. In terms of a minimal basis of on-shell kinematics, the ansatz  starts with 45 free parameters. We enforce color-kinematics by imposing antisymmetry of each vertex as well as  kinematic Jacobi on every edge of both graphs. All remaining parameters  are constrained uniquely via the four-pion cut \eqn{fourPionCut} and the vector Ward identity.
In contrast to the amplitudes constructed from the $\mathcal{L}^{\cP}$ of \eqn{gaugeNLSMaction}, we can now build the partial amplitudes that satisfy the color-dual BCJ relations at five-point,
\begin{align}
\label{eq:dcc5point}
\begin{split}
 A_5(1_\pi,2_\pi,3_\pi,4_\pi,5_A)& =g\Lambda\Bigg[\frac{ s_{35} \kappa_2^{(5)} -s_{25} \kappa_3^{(5)}}{s_{23}}
 +\frac{ s_{35} s_{25} \kappa_{12}^{(5)} }{s_{12} s_{34}}
+\frac{ s_{25} \kappa_3^{(5)}}{s_{34}}-\frac{ s_{35} \kappa_2^{(5)}}{s_{12}}\\
  & \hspace{1.8in}  +3\left(\frac{ s_{24} \kappa_1^{(5)}}{s_{15}}-\frac{ s_{13} \kappa_4^{(5)}}{s_{45}}+\kappa_{24}^{(5)}\right)\Bigg],
   \end{split}
\end{align}
where  $s_{ij}=(k_i + k_j)^2$, and   $\kappa_{i \dots j}^{(5)} \equiv ( k_i + \dots  + k_j) \cdot \varepsilon_5$. 

Now we can see what was missing from the simple gauged NLSM in order to manifest the duality between color and kinematics. Had we just considered the contribution from $\mathcal{L}^{\cP}$, only the final term in \eqn{eq:dcc5point} would have been generated. As can be seen in the first line, the newly constructed $A(1_\pi,2_\pi,3_\pi,4_\pi,5_A)$ has a two-propagator residue, admitting a non-vanishing maximal-cut proportional to $g\Lambda $ contributing to the secondary $g_2$ topology,
\[
\mathcal{A}_5(1_\pi,2_\pi,3_\pi,4_\pi,5_A)|_{g_2\text{ Max.Cut}}=	\fivegraphMcut{3}{4}{5}{1}{2} \, .
\]
If we had required that the above channels vanish (as they do for the theory of \eqn{gaugeNLSMaction}) then we would not have found a color-dual solution.  
  This cut can only be satisfied at this mass-dimension (in a non-abelian manner) by the inclusion of  
  amplitudes associated with the $\Tr(F^3)$ operator\footnote{Additional higher-derivative contact interactions may be required by double-copy consistency. {We also note that the $\Tr(F^3)$ operator that we consider here is the only other independent all-vector three-point amplitude besides the one contained in pure YM.  In our notation, it is given explicitly by $\Tr (F^3) = \Tr( F_\mu{}^\nu F_\nu{}^\rho F_\rho{}^\mu )$.}} 
whose coupling is fixed to be $g_{F^3}= \Lambda/g$.  Intriguingly, $\Tr(F^3)$ was the first higher-derivative vector operator found to be compatible~\cite{Broedel2012rc} with adjoint-type double-copy, at least for all amplitudes admitting only single-insertions~\cite{SchlottererBGCurrent}. {This leads us to the following starting point for the double-copy consistent theory for YM + $\pi$,
\be
\mathcal{L}^{\rm dcc}_{\text{YM}+ \pi} \supset \mathcal{L}^{\text{cov.} \pi} - \frac{g_{F^3}}{3 } \Tr ( F^3) + \dots  \ .
\ee 
 In the next section we will demonstrate that this is just the beginning. Once $\Tr(F^3)$ forces itself into the conversation, we find for double copy consistency that an infinite tower of scalar-vector operators are needed at fixed multiplicity, thereby pushing both scalar and vector sectors of the theory into concordance with particular ultraviolet completions.

%
%
%

\section{Completing the vector-pion theory under double-copy consistency}\label{sect:bootstrapF3Tower}

  %
  %
  %
\subsection{New contributions for vector amplitudes with zero scalars}
The first set of amplitudes to consider are the all-vector observables resulting from the inclusion of $\Tr(F^3)$ to the Lagrangian of \eqn{gaugeNLSMaction}. As reviewed in \sect{sect:vectorTower}, these contributions were identified in~\cite{f3Ladder}, where the interplay between pure Yang-Mills theory and $\Tr(F^3)$ induces a ladder of higher-derivative four-point local contact terms. The coefficients for a subset of the available color-dual contacts are completely fixed by the gauge coupling, $g$, and the dimensionful $\Tr(F^3)$ coefficient, $g_{F^3}= \Lambda/g$, when we demand that the theory is double-copy consistent. By setting any additional higher-derivative freedom to zero, we find that for $k\geq 2$, the required color-dual contacts take the form:
\begin{equation}
\label{vectorContacts}
 A_4^{(k)} (1_A, 2_A,  3_A, 4_A) = g^2 \left(\frac{\Lambda}{g^2}\right)^{k }    u 
\times \Bigg[\frac{\tr[F_1F_2]\tr[F_3F_4]}{s_{12}^2}s_{12}^{k-1}+\text{cyc}(2,3,4)\Bigg],
\end{equation}
where 
\begin{equation}
\label{fTensorDef}
F_i^{\mu\nu} = k_i^\mu {\varepsilon}_i^\nu- k_i^\nu \varepsilon_i^\mu\,, 
\end{equation}
and 
\begin{equation}
 A^{(F^3)^2+F^4}_4 \equiv   A_4^{(2)}, \qquad  A_4^{D^{2k}F^4} \equiv A^{(k+2)}_4.
\end{equation}
It will be helpful to note that $A^{(1)}_4$ is intimately related to the $A^{(DF)^2}$ amplitude of \cite{Johansson:2017srf} as written in $D$-dimensions~\cite{Azevedo2018dgo},
\begin{equation}
A^{(1)}_4=A^{F^3}_4-A^{(DF)^2}_4\,.
\end{equation}
As stated above, these are precisely the building blocks that appear in the inductive ladder described in \sect{sect:vectorTower}.  By considering pure-vector contacts that contribute to two-scalar three-vector five-point amplitudes via factorization, we have now been able to explicitly verify \eqn{vectorContacts} through $\mathcal{O}(\Lambda^5)$. 

In ref.~\cite{f3Ladder} it was conjectured that the resulting two-parameter effective field theory can be resummed to $(DF)^2+\text{YM}$ theory, first identified in \cite{Johansson:2017srf} when considering whether conformal gravity is double-copy constructible. The amplitudes generated by the $(DF)^2+\text{YM}$ Lagrangian have been used in the double-copy construction of open bosonic string and heterotic string amplitudes at tree-level. The conjectured resummation of the inductive ladder in ref.~\cite{f3Ladder} is due to the observation that if one collects $A_4^{(k)}$ to all orders in mass-dimension, this is equivalent to a low energy expansion of the four-point ordered $(DF)^2+\text{YM}$ amplitude:
\begin{equation}
\label{lowEnergyB}
\begin{split}
g^{-2}A^{(DF)^2+\text{YM}}(1234) &= A^{\text{YM}}_{(1234)}+\alpha' A^{(DF)^2}_{(1234)}+ \alpha' u\Bigg[\frac{\tr[F_1F_2]\tr[F_3F_4]}{s^2_{12}(1+\alpha' s_{12})}+\text{cyc}(2,3,4)\Bigg]
\\
&=A^{\text{YM}}_{(1234)}+\alpha' A^{F^3}_{(1234)} + \sum_{k=2}^\infty \alpha'^{\,k} A^{(k)}_{(1234)}\,.
\end{split}
\end{equation}
Recall, as stated in \sect{sect:vectorTower}, the dimensionful coupling $\alpha' = -\Lambda/g^2$. The additional freedom alluded to above was then described in ref.~\cite{f3Ladder} by the following double-copy consistent higher-derivative vector amplitude:
\begin{equation}
\label{eq:dccVectorAmp}
A^{\text{dcc}}_{(1234)} = A^{(DF)^2+\text{YM}}(1234)\left[1+\sum_{x\geq 1, y} c_{(x,y)}\sigma_3^x \sigma_2^y\right]
\end{equation}
where $\sigma_3 =(stu)/8$ and $\sigma_3 =(s^2+t^2+u^2)/8$ are permutation invariants of four-point kinematics. Again, by considering two-scalar three-vector factorization, we have verified that the double-copy consistent building blocks in our tower are reproduced by the low energy expansion of \eqn{eq:dccVectorAmp} up through $\mathcal{O}(\Lambda^5)$, at which point there are two additional free parameters, $c_{(1,0)}$ and $c_{(1,1)}$. Thus, it would appear that in order for our candidate vector-pion theory to satisfy the duality between color and kinematics to all-multiplicity, we are interested in a Lagrangian of the form, $(DF)^2+\text{YM}+\pi +\ldots$, where the ellipses denote the potential of additional higher-derivative operators that are unconstrained by double-copy consistency. 

For the remainder of this section, we will focus our attention on the operators whose Wilson coefficients are fixed completely in terms of $\Lambda$ and $g$, and thus required by double-copy consistency, and allow all unconstrained freedom in the free parameters, $c_{(x,y)}$, to vanish.

  %
  %
  %
\subsection{New contributions for vector amplitudes with two scalars}

Now, in our vector-scalar theory, the inclusion of $\Tr(F^3)$ creates a new $\mathcal{O}(\Lambda)$ contribution at four points to the two-pion, two-vector scattering amplitude, $\mathcal{A}_4(\pi \pi AA)$.  Consider the four-point scattering where legs one and two are pions, and legs three and four are vectors.  Demanding factorization to the correct $\Tr(F^3)$ vertex and double-copy consistency,  there is a new $\Lambda$ contribution to the partial amplitude,  
  \begin{equation}
  \label{2pi2glueLambda}
     A_4^{(1)} (1_\pi , 2_\pi, 3_A, 4_A ) =  \, {  \Lambda} \frac{u}{s} \Big[(k_3\cdot \varepsilon_4)(k_4\cdot \varepsilon_3)-(k_3\cdot k_4) (\varepsilon_3\cdot\varepsilon_4)\Big] \,,
\end{equation}
where $s=(k_1+k_2)^2$,  $u=(k_1+k_3)^2=-s-t$, and for the remainder of the paper, the superscript ${}^{(k)}$ denotes $k$-th order in $\Lambda$.  

We also find that demanding factorization and double-copy consistency has forced an additional two-pion-two-vector contact operator.  This can most easily be seen by considering the other ordered amplitudes given by the BCJ relations.  At four-point these are  
\begin{equation} \label{BCJ4point}
	\frac{A_4 (1_\pi , 2_\pi, 3_A, 4_A ) }{u}= \frac{A_4  (1_\pi , 2_\pi, 4_A, 3_A ) }{t}=\frac{A_4  (1_\pi , 3_A, 2_\pi,  4_A ) }{s} \ , 
\end{equation}
which is true for our amplitudes by construction.  Then, we can see from \eqn{2pi2glueLambda} that the $s$-channel pole in $A_4  (1_\pi , 3_A, 2_\pi,  4_A )  $ cancels, making it a pure contact term.  
For completeness, the full color-dressed amplitude is given by
\begin{equation}
    \mathcal{A}_4 ( \pi \pi AA)= \left(c_s u  + c_u s \right ) \frac{A_4 (1_\pi , 2_\pi, 3_A, 4_A ) }{u} .
\end{equation}

\begin{figure*}[t]
    \centering
    \begin{equation*}
\begin{split}
A^{(n)}{(1^\pi 2^A 3^\pi 4^A5^A)}\Big|_{s_{45}-\text{cut}}=\fivegraphScalarPropOp{1}{2}{4}{5}{\Lambda^{n-1}}{\Lambda^1}{3}{hred}{hgrey1}+ \fivegraphScalarPropOp{1}{2}{4}{5}{\Lambda^n}{\Lambda^0}{3}{hblue}{hgrey2} \qquad {\mathcal{O}(\Lambda^n)}
\end{split}
\end{equation*}
   \begin{equation*}
\begin{split}
A^{(n+1)}{(1^\pi 2^A 3^\pi 4^A5^A)}\Big|_{s_{45}-\text{cut}}=\fivegraphScalarPropOp{1}{2}{4}{5}{\Lambda^{n}}{\Lambda^1}{3}{hred}{hgrey2}+ \fivegraphScalarPropOp{1}{2}{4}{5}{\Lambda^{n+1}}{\Lambda^0}{3}{hblue}{hgrey3} \qquad {\mathcal{O}(\Lambda^{n+1})}
\end{split}
\end{equation*}
    \caption{\label{fig:inductiveLadder2pi2glue}Inductive ladder of higher-derivative contacts generated by consideration of color-dual consistency and cuts of the form $A(\pi\pi AA)\times (A_3^\text{YM}+A_3^{F^3})$. }  
\end{figure*}

With the inclusion of $\Tr(F^3)$ and an $\mathcal{O}(\Lambda)$ contribution to $A_4 ( \pi \pi AA)$ required by double-copy consistency, we are now forced to consider contributions to $\mathcal{A}_5(\pi\pi AAA)$ at $\mathcal{O}(\Lambda^2)$.    We proceed with the same procedure as the previous section, constructing an ansatz that respects mass-dimension and little-group scaling, and then impose BCJ relations, and we will find an inductive process, in the same spirit to that identified in~\cite{f3Ladder} and reviewed in \sect{sect:vectorTower}, that generates a tower of higher derivative operators with Wilson coefficients fixed in terms of $\Lambda$ and $g$.

After requiring that our kinematic weights are color-dual, we can fix the remaining freedom on consistent factorization to the available lower multiplicity building blocks. At this mass-dimension, the $\mathcal{O}(\Lambda)$ partial amplitude of \eqn{2pi2glueLambda} will contribute to a three-to-two-particle generalized unitarity cut with $\Tr(F^3)$. This will leave the $\Tr(F^2)$ contribution to the two-particle cut available to interact with a potential $\mathcal{O}(\Lambda^2)$ contact on the other side of the cut, if required by double-copy consistency (shown in \Fig{fig:inductiveLadder2pi2glue} for general mass-dimension). Indeed, we find that an additional four-point contact term is required of the form, 
  \begin{equation}
  \label{2pi2glueLambda2}
     A_4^{(2)}  (1_\pi , 2_\pi, 3_A, 4_A ) =     \frac{  \Lambda^2 }{g^2 }  \, u \Big[(k_3\cdot \varepsilon_4)(k_4\cdot \varepsilon_3)-(k_3\cdot k_4) (\varepsilon_3\cdot\varepsilon_4)\Big] \,,
\end{equation}
for our candidate vector pion theory to continue to respect double-copy consistency. By the same logic, this now requires us to further consider contributions at the next order in $\Lambda$ -- yielding yet another four-point contact at $\mathcal{O}(\Lambda^3)$. In fact, we observe that in order for $\mathcal{A}_5(\pi\pi AAA)$ to be consistent to all orders in $\Lambda$, contact terms contributing to the partial amplitude, $A_4^{(k)}(\pi\pi AA)$, at $\mathcal{O}(\Lambda^k)$ must be completed by an additional contact $A_4^{(k+1)}(\pi\pi AA)$ at $\mathcal{O}(\Lambda^{k+1})$, as required by double-copy consistency. This inductive process is sketched in \Fig{fig:inductiveLadder2pi2glue}. It turns out that the general form of these color-dual four-point contributions at each additional order in mass-dimension can evidently be written as,
\begin{equation}
\label{2pi2glue}
A_4^{(k)} (1_\pi , 2_\pi, 3_A, 4_A ) =  g^2 \left(\frac{\Lambda}{g^2}\right)^{k }   u\, s^{k-1} \frac{ \tr [F_3 F_{4} ]}{s}\,,
\end{equation} 
where the superscript $k$ counts additional orders in mass-dimension, and $F_i^{\mu\nu}$ is given in \eqn{fTensorDef}.  The resulting tensor, 
\be
  \tr [ F_i F_j  ] \equiv F_i^{\mu \nu} F_j^{\rho \sigma} \eta_{\mu \rho} \eta_{\nu \sigma} = - 2 \Big[(k_i \cdot \varepsilon_j)(k_j \cdot \varepsilon_i)-(k_i \cdot k_j ) (\varepsilon_i \cdot\varepsilon_j )\Big]  \ ,
\ee
is the same tensor structure found in \eqn{2pi2glueLambda} and \eqn{2pi2glueLambda2}, for example. We have explicitly verified \eqn{2pi2glue} up to $\mathcal{O} ( \Lambda^5)$ when unconstrained freedom in higher-derivative Wilson coefficients is set to zero.

%
%
\subsection{New contributions for vector amplitudes with four scalars}
\begin{figure}[t]
    \centering
    \begin{equation}
\begin{split}
A{(1^\pi 2^\pi 3^\pi 4^\pi 5^A)}\supset \fivegraphMostScalarIntGluePropOp{3}{4}{1}{2}{\Lambda^{n}}{\Lambda^0}{5}{hblue}
+
\fivegraphMostScalarExtGluePropOp{1}{2}{4}{5}{\Lambda^{n}}{\Lambda^0}{3}{hblue} 
\end{split}
\end{equation}

    \caption{\label{fig:4pi1glue}Generalized unitarity cuts used to constrain BCJ satisfying partial amplitudes contributing to $\mathcal{A}_5(\pi\pi\pi\pi A)$. Given the existence of $A_4^{(n)}(\pi\pi AA)$ at any order $\mathcal{O}(\Lambda^n)$, an additional four-point all-pion higher-derivative contact is required to maintain double-copy consistency at five-point, such that $A_4^{(n)}(\pi\pi AA) \iff A_4^{(n)}(\pi\pi \pi\pi)$. } 
\end{figure}

With this tower of contacts contributing to $\mathcal{A}_4(\pi\pi AA)$ in hand, we can immediately see that additional higher derivative corrections must be added to the all-pion amplitude, $\mathcal{A}_4(\pi\pi\pi\pi)$, in order for $\mathcal{A}_5(\pi\pi \pi \pi A)$ to be double-copy consistent to all orders in mass-dimension.

Imposing the factorization constraints in \Fig{fig:4pi1glue}, and setting any remaining freedom to zero, we find that the four-point ordered amplitude contributing to the all-pion cuts at a given order in $\Lambda$ takes the form,
\begin{equation}
\label{allpion}
A_4^{(k)} (1_\pi, 2_\pi,  3_\pi, 4_\pi) = g^2 \left(\frac{\Lambda}{g^2}\right)^{k }  u \,  \Big[s^{k-1}+ t^{k-1} + u^{k-1} \Big] \ . 
\end{equation}
We have verified through $\mathcal{O}(\Lambda^5)$ that four-point amplitudes of the form stated in \eqn{2pi2glue} and \eqn{allpion} are required for double-copy consistency at five points. 

\begin{table}[b]
\begin{center}

\begin{tabular}{||c||*{5}{c|}}\hline

\diagbox{$\mathcal{O}(\Lambda^n)$}{$||\pi||=2k$}
&\makebox[3em]{$k=0$}&\makebox[3em]{$k=1$}\\\hline\hline
$n=0$ &\cellcolor{hgreen1}\threeTableGraph{gluon}{gluon}{gluon}{\text{YM}}{hblue}&\cellcolor{hgreen1}\threeTableGraph{}{}{gluon}{\text{YMS}}{hgrey0}\\\hline
$n=1$ &\cellcolor{hblue1}\threeTableGraph{gluon}{gluon}{gluon}{F^3}{hred}&\\\hline

\end{tabular}
\caption{Three-point amplitudes for $2k$ pions -- Green cells contain amplitudes generated by \eqn{gaugeNLSMaction}. The additional $\Tr(F^3)$ interaction highlighted in yellow was needed in \eqn{eq:dcc5point} for double-copy consistency.}
\label{tab:3point}
\end{center}
\end{table}

\section{Analysis of the double-copy consistent vector-pion theory}\label{sect:novelFeaturesF3}
In this section we will identify out some novel features of this candidate vector-pion theory, and some directions for future studies. To begin, we provide a set of tables that display the additional color-dual operators that are required for our conjectured vector-pion theory. All cells that appear with a diagram indicate that the amplitude has been constructed, and is available in the ancillary file. 

%
%
\subsection{Summary of new contributions}\label{sect:tables}

\begin{table}[t]
\begin{center}

\begin{tabular}{||c||*{6}{c|}}\hline

\diagbox{$\mathcal{O}(\Lambda^n)$}{$||\pi||=2k$}
&\makebox[3em]{$k=0$}&\makebox[3em]{$k=1$}&\makebox[3em]{$k=2$}\\\hline\hline
$n=0$ &\cellcolor{hgreen1}\fourTableGraph{gluon}{gluon}{gluon}{gluon}{\text{YM}}{hgrey0}\textbf{0}&\cellcolor{hgreen1}\cellcolor{hgreen1}\fourTableGraph{gluon}{gluon}{}{}{\text{YMS}}{hgrey0}\textbf{0}&\cellcolor{hgreen1}\cellcolor{hgreen1}\fourTableGraph{}{}{}{}{\text{YMS}}{hgrey0}\textbf{0}\\\hline

$n=1$&\cellcolor{hblue1}\fourTableGraph{gluon}{gluon}{gluon}{gluon}{F^3}{hgrey1}\textbf{0}&\cellcolor{hblue1}\fourTableGraph{gluon}{gluon}{}{}{}{hgrey1}\textbf{0}&\cellcolor{hgreen1}\fourTableGraph{}{}{}{}{\text{NLSM}}{hgrey1}\textbf{0}\\\hline

$n=2$&\fourTableGraph{gluon}{gluon}{gluon}{gluon}{F^4}{hgrey2}\textbf{0}&\fourTableGraph{gluon}{gluon}{}{}{}{hgrey2}\textbf{0}&\fourTableGraph{}{}{}{}{}{hgrey2}\textbf{0}\\\hline

$n=3$&\fourTableGraph{gluon}{gluon}{gluon}{gluon}{D^2F^4}{hgrey3}\textbf{1}&\fourTableGraph{gluon}{gluon}{}{}{}{hgrey3}\textbf{1}&\fourTableGraph{}{}{}{}{}{hgrey3}\textbf{1}\\\hline

$n=4$&\fourTableGraph{gluon}{gluon}{gluon}{gluon}{D^4F^4}{hgrey4}\textbf{1}&\fourTableGraph{gluon}{gluon}{}{}{}{hgrey4}\textbf{1}&\fourTableGraph{}{}{}{}{}{hgrey4}\textbf{1}\\\hline

$n=5$&\fourTableGraph{gluon}{gluon}{gluon}{gluon}{D^6F^4}{hgrey5}\textbf{2}&\fourTableGraph{gluon}{gluon}{}{}{}{hgrey5}\textbf{2}&\fourTableGraph{}{}{}{}{}{hgrey5}\textbf{2}\\\hline

\end{tabular}
\caption{Four-point amplitudes for $k$ pairs of pions -- Green cells contain amplitudes generated by \eqn{gaugeNLSMaction}. The additional interactions highlighted in yellow were needed in \eqn{eq:dcc5point} for double-copy consistency.}
\label{tab:4point}
\end{center}
\end{table}
\begin{table}[t]
\begin{center}

\begin{tabular}{||c||*{6}{c|}}\hline

\diagbox{$\mathcal{O}(\Lambda^n)$}{$||\pi||=2k$}
&\makebox[3em]{$k=0$}&\makebox[3em]{$k=1$}&\makebox[3em]{$k=2$}\\\hline\hline

$n=0$ &\cellcolor{hgreen1}\fiveTableGraph{gluon}{gluon}{gluon}{gluon}{gluon}{\text{YM}}{hgrey0}\textbf{0}&\cellcolor{hgreen1}\cellcolor{hgreen1}\fiveTableGraph{gluon}{gluon}{}{}{gluon}{\text{YMS}}{hgrey0}\textbf{0}&\cellcolor{hgreen1}\cellcolor{hgreen1}\fiveTableGraph{}{}{}{}{gluon}{\text{YMS}}{hgrey0}\textbf{0}\\\hline

$n=1$&\fiveTableGraph{gluon}{gluon}{gluon}{gluon}{gluon}{}{hgrey1}\textbf{0}&\fiveTableGraph{gluon}{gluon}{}{}{gluon}{}{hgrey1}\textbf{0}&\cellcolor{hred1}\fiveTableGraph{}{}{}{}{gluon}{\text{cov.}\pi}{hgrey1}\textbf{0}\\\hline

$n=2$&\fiveTableGraph{gluon}{gluon}{gluon}{gluon}{gluon}{}{hgrey2}\textbf{0}&\fiveTableGraph{gluon}{gluon}{}{}{gluon}{}{hgrey2}\textbf{0}&\fiveTableGraph{}{}{}{}{gluon}{}{hgrey2}\textbf{0}\\\hline

$n=3$&\fiveTableGraph{gluon}{gluon}{gluon}{gluon}{gluon}{}{hgrey3}\textbf{1}&\fiveTableGraph{gluon}{gluon}{}{}{gluon}{}{hgrey3}\textbf{1}&\fiveTableGraph{}{}{}{}{gluon}{}{hgrey3}\textbf{1}\\\hline

$n=4$&\fiveTableGraph{gluon}{gluon}{gluon}{gluon}{gluon}{}{hgrey4}\textbf{1}&\fiveTableGraph{gluon}{gluon}{}{}{gluon}{}{hgrey4}\textbf{1}&\fiveTableGraph{}{}{}{}{gluon}{}{hgrey4}\textbf{1}\\\hline

$n=5$&&&\fiveTableGraph{}{}{}{}{gluon}{}{hgrey5}\textbf{3}\\\hline

\end{tabular}
\caption{Five-point amplitudes for $k$ pairs of pions -- Green cells contain amplitudes generated by \eqn{gaugeNLSMaction}. The amplitude highlighted in red violated BCJ relations when only considering interactions of \eqn{gaugeNLSMaction}. This amplitude, $A^{\text{cov.}\pi}$, needed additional interactions in \eqn{eq:dcc5point} to be double-copy consistent.}
\label{tab:5point}
\end{center}
\end{table}

Cells in green correspond to amplitudes that are generated by the Feynman rules of \eqn{gaugeNLSMaction}, our starting point. The four-pion five-point amplitude in Table \ref{tab:5point} is highlighted in red, to indicate that it is the first amplitude that forced us to consider additional operators; these additional operators, that served as the base-steps of our inductive tower, are highlighted in yellow in Tables \ref{tab:3point} and \ref{tab:4point}. Finally, the bold-face numbers in each cell correspond to the number of free Wilson-coefficients generated by our bootstrap. These additional parameters are unconstrained by double-copy consistency and five-point factorization; however there could in principle be additional constraints at higher multiplicity.

As outlined in the review of unitarity methods in \sect{sec:unitarityMethods}, these amplitudes were constrained by (1) color-kinematics (2) factorization and (3) gauge invariance. To elucidate the requisite four-point contacts through $\mathcal{O}(\Lambda^5)$ and the available freedom, we only considered color-kinematic and factorization constraints on the same mass-dimension five-point amplitude; and did not perform the computationally onerous process of further imposing gauge-invariance. For this reason, the zero- and two-scalar five-point amplitudes at $\mathcal{O}(\Lambda^5)$ are left blank.

In addition to the three-, four- and five-point amplitudes, we have constructed the all-scalar six-point amplitude at NLSM mass-dimension, $\mathcal{O}(\Lambda^2)$. Due to the inclusion of a five-point amplitude at $\mathcal{O}(\Lambda^2 /g)$ required by double-copy consistency, there is a novel correction to the pure NLSM six-point amplitude, shown in Table \ref{tab:6point}. We will discuss the implications of this in more detail in the next subsection. 

\begin{table}[t]
\begin{center}

\begin{tabular}{||c||*{5}{c|}}\hline

\diagbox{$\mathcal{O}(\Lambda^n)$}{$||\pi||=2k$}
& \multicolumn{2}{c|}{\makebox[3em]{$k=3$}}\\\hline\hline
$n=2$ &\cellcolor{hgreen1}{{\sixTableGraph{}{}{}{}{}{}{hgrey}{\text{NLSM}}}}&+\sixTableGraph{}{}{}{}{}{}{hgrey}{A^\text{new}_6}\\\hline

\end{tabular}
\caption{Six-point amplitude for 3 pairs of pions -- Green cells contain amplitudes generated by \eqn{gaugeNLSMaction}. The additional contribution at $\mathcal{O}(\Lambda^2)$ is required for double-copy consistency of the full vector-pion theory.}
\label{tab:6point}
\end{center}
\end{table}
%
%
%
\subsection{Modifying NLSM and thus DBIVA behavior}\label{sect:NLSMmod}
As suggested in Table \ref{tab:6point}, the existence of five-point contributions at $\mathcal{O}(\Lambda^2/g)$ to $\mathcal{A}(\pi\pi\pi\pi A)$ mandates corrections to the already color-dual six-point pion amplitude for our theory. This additional $\mathcal{O}(\Lambda^2/g)$ term leads to non-trivial corrections to the pure pion six-point amplitude at $\mathcal{O}(\Lambda^2)$, such that 
\begin{equation}
\label{6pointDecomp}
A_6^{(DF)^2+\text{YM}+\pi} (123456)\Big|_{\Lambda^2} = A_6^{\pi} (123456)
+A_6^{\text{new} }(123456)
\end{equation}
where the first term is the usual six-point NLSM partial amplitude:
\begin{equation}
A_6^{\pi} (123456) = 9\Lambda^2 \left[\frac{s_{13}s_{46}}{s_{123}}+\frac{s_{24}s_{51}}{s_{234}}+\frac{s_{35}s_{62}}{s_{345}} - s_{135}\right]
\end{equation}
and the novel contribution appearing at $\mathcal{O}(\Lambda^2)$ takes the form:
\begin{multline}
A_6^{\text{new} }(123456) = \Lambda^2 \Bigg[\frac{1}{2}\left( \frac{(s_{15}+s_{25})\tau_{12|34}}{s_{12}s_{34}}+\frac{s_{36}\tau_{12|45}}{s_{12}s_{45}}+\frac{(s_{35}+s_{36})\tau_{12|56}}{s_{12}s_{56}}\right)
\\
+ \frac{\tau_{12|56}-\tau_{12|45}}{s_{12}}- (s_{14}+s_{23}) + \text{cyc}(1,2,3,4,5,6)\Bigg]
\end{multline}
where we have introduced the notation, $\tau_{ij|kl}=s_{ik}s_{jl}-s_{il}s_{jk}$.   The novel contribution, $A^{\text{new}}_6$, has non-vanishing residues on cuts that expose individual three-point amplitudes,
\begin{equation}
\sixTableGraph{}{}{}{}{}{}{hgrey}{A_6|_{\Lambda^2}\,} \supset \cutDiagramSix\,.
\end{equation}
This new contribution alters the Addler's zero behavior of the theory.  Note that  $A^{\text{new}}_6$  does not contribute to typical factorization channel available to six-point NLSM pion amplitudes, the sewing of two four-point trees. It is particularly interesting to note that this novel contribution to the six-point amplitude independently satisfies the BCJ relations, unlike the additional term in \eqn{eq:dcc5point} that was required to make $\mathcal{A}(\pi\pi\pi\pi A)$ color-dual.  The identification of the coupling with $\Lambda^2$ is only required by double-copy consistency with the lower-multiplicity results found thus far. The gauge theory cut, depicted above, will then lead to a non-vanishing factorization channel in the resulting double-copy to $\text{DBIVA}+$Weyl-Einstein supergravity:
\begin{equation} 
\mathcal{A}_6(\lambda^{\text{VA}}\cdots\lambda^{\text{VA}})\Big|_{l^2\text{-cut}} \!\!\!\! \sim\sum_{ \text{states}} \mathcal{A}_5(\lambda^{\text{VA}}\cdots\lambda^{\text{VA}}, l^{\bar{s}})\mathcal{A}_3(-l^{s}, \lambda^{\text{VA}}\lambda^{\text{VA}})
\end{equation}
This is obviously in contrast to the  usual behavior of pure $\text{DBIVA}$, that mandates factorization to three-point trees vanish in the scattering. As both contributions above in \eqn{6pointDecomp} appear at $\mathcal{O}(\Lambda^2)$, it is not possible to tune the interaction to recover pure DBIVA theory from the resulting double-copy. 

We must admit this is a somewhat disturbing aspect of our construction -- we wished to build Volkov-Akulov coupled to supergravity, but we seem to have altered the very soft-limit behavior we identify with Volkov-Akulov scattering to begin with. Does this destroy the nilpotence associated with the fermionic field in DBIVA uncoupled to supergravity?  The answer to that question awaits future work. We note that the more radical construction presented in \sec{sect:radicalModel} does leave DBIVA behavior unmodified to all multiplicity at the cost of manifest exchange symmetry between scalars in the constituent gauged NLSM theory.
 
%
%
%

%
%
%
\subsection{Resumming to dimensionally reduced $(DF)^2+\text{YM}$}

Was it in some sense inevitable that an effective field theory of  identical scalars, NLSM, could only be combined with an effective field theory of vectors, $(DF)^2+\text{YM}$, in a double-copy consistent way? We are prepared to conjecture an answer by considering the suggestive field theory question:  what is the dimensional reduction of  $(d+1)$-dimensional  $(DF)^2+\text{YM} $, a known color dual theory, to $d$-dimensions? 

We can perform a dimensional reduction of this form directly at the level of amplitudes. To do so, we will define the $(d+1)$-dimensional momenta, $\mathcal{K}_a$, and polarizations, $\mathcal{E}_a$, as follows:
\begin{equation}
\mathcal{K}_a = (k_\mu ,0)_a\qquad \mathcal{E}_a = \begin{cases} 1& a = d
\\
\varepsilon_\mu & a \neq d
\end{cases}
\end{equation}
where $\mu$ takes on values in the lower dimensional subspace, $\mu = 0,...,d-1$. When we identify the pions with the extra-dimensional polarization mode, $\mathcal{E}_{a=d}$, one finds that the building blocks in \eqn{lowEnergyB} exactly reproduce the scalar amplitudes in  \eqn{2pi2glue} and \eqn{allpion}. The same goes the for available five-point amplitudes through $\mathcal{O}(\Lambda^4)$. 

Furthermore, this dimensional reduction procedure is in one-to-one correspondence with the application of transmutation operators, $\mathcal{T}_{ij}=\frac{\partial}{\partial (\varepsilon_i\varepsilon_j)}$, of Cheung, Shen, and Wen~\cite{Cheung:2017ems,Cheung:2017yef}, related to the `generalized dimensional reductions' of Cachazo, He, and Yuan~\cite{Cachazo2014xea}. So the observation that the all-vector building blocks  in \eqn{vectorContacts}, generated by $\mathcal{L}_{(DF)^2+\text{YM}}$, are secretly related to the mixed vector-pion color-dual building blocks identified in \eqn{2pi2glue} and \eqn{allpion} can be neatly captured by the following set of relations:
\begin{equation}
A_4^{(n)}(1_\pi,2_\pi,3_A,4_A) = \mathcal{T}_{12} \circ A^{({n})}_4(1_A,2_A,3_A,4_A)
\end{equation}
\begin{equation}
A_4^{(n)}(1_\pi,2_\pi,3_\pi,4_\pi) =\left(\mathcal{T}_{12}\mathcal{T}_{34}+\text{cyc}(2,3,4)\right)\circ A^{({n})}_4(1_A,2_A,3_A,4_A)
\end{equation}
Dimensional reduction does not seem to affect the double-copy consistency \cite{BCJreview}, but does offer consistently interacting scalars.  Likewise, performing this dimensional reduction on the ${(DF)^2+\text{YM}}$ Lagrangian yields the type of interaction terms required to build the color-dual mixed amplitudes we have presented here through six-points. So adding appropriately weighted higher derivative vector operators to the covariantized NLSM seems exactly the right approach to build a theory compatible with the double-copy web of theories through the interactions we have yet explored. 

The double-copy of sYM with ${(DF)^2+\text{YM}}$ is known to yield a Weyl-Einstein type conformal supergravity (CSG) \cite{Johansson:2017srf}, 
\begin{equation}
A^{\text{CSG}} = A^{\text{sYM}}\otimes A^{(DF)^2+\text{YM}}\,.
\end{equation}
This theory is related to the gravitational amplitudes in the heterotic string by a particular choice of Wilson coefficients for an additional tower of  double-copy consistent higher derivative operators.  Our conjectured vector-pion theory double-copied with sYM is then related to this family of Weyl-Einstein type conformal supergravities, by what appears to be a sort of dimensional reduction -- but one that only applies to one factor in the double-copy construction. Our model appears consistent with the behavior of modified Volkov-Akulov fermions dynamically coupled to spacetime of interest in $\alpha$-attractor models, which naturally reside in gravity theories where the super \Poincare symmetry is upgraded to the super-conformal group \cite{Kallosh:2013yoa}.  This arises from a candidate vector-pion theory that is related to the color-dual $(DF)^2+\text{YM}$ via dimensional-reduction. All-order in $\Lambda$ predictions in this theory could be extracted directly from any multiplicity $(DF)^2+\text{YM}$ theory amplitudes. 

Moreover, constructing the Lagrangian for our vector-pion theory, with all residual higher-derivative freedom set to zero, is just a matter of redefining the gauge fields, $A_a \rightarrow (A_\mu ,\pi)_a$ and projecting the derivatives down to the lower dimensional subspace, $\partial_a \rightarrow (\partial_\mu ,0)_a$. This gives us the following conjecture for the Lagrangian that reproduces the amplitudes constructed in this paper:

\begin{equation}
\mathcal{L}_{(DF)^2+\text{YM}+\pi} \equiv \mathcal{L}_{(DF)^2+\text{YM}} \Big|^{\partial_a \rightarrow (\partial_\mu ,0)_a}_{A_a \rightarrow (A_\mu ,\pi)_a}\,.
\end{equation}

The observation that a double-copy consistent vector-pion theory could be equivalent to the dimensional reduction of $(DF)^2+\text{YM}+\text{HD}$ also sheds some light on the question of why we cannot recover the usual NLSM Adler's zero condition from taking limits of coupling in our theory.   Since the $\alpha'$ in $(DF)^2+\text{YM}$-amplitudes is related to the pion decay-width, $\alpha' = -\Lambda/g^2$, the limit $g\rightarrow 0$ is equivalent to the $\alpha'\rightarrow \infty$ limit, where the theory becomes a dimensional reduction of the dimension six theory, $\mathcal{L}^{(DF)^2}$\cite{Johansson:2017srf}. In this limit, dimensional reduction is one-to-one with dimensional truncation and all pionic amplitudes vanish. 
%
%
%
\subsection{Helicity selection rule from new contributions}

The tensor structure of \eqn{2pi2glueLambda} is different from the standard $\Tr ( F^2 )$ contribution, which has particularly sharp implications for the helicity structure of gluon scattering in four dimensions.  The standard $\Tr(F^2)$ term only contributes to this amplitude for mixed-helicity gluons, while the $\Tr(F^3)$ operator only contributes to this amplitude for same-helicity gluons.
Using the four-dimensional spinor-helicity formalism, when the gluons are both plus helicity the amplitude is simply 
\begin{equation}
   A_4(1_\pi,2_\pi,3_A^+,4_A^+) =  \Lambda \frac{u}{s} {\left[3 4\right]}^2 \ ,
\end{equation}
which starts at $\mathcal{O}(\Lambda)$ because the standard YM vertex does not contribute.  
It is worth spending a few words considering the implications this has for the double-copy theory we are attempting to build.  Aiming for $\text{DBIVA}+\text{supergravity}$, double-copying these amplitudes with $\mathcal{N}\leq4$ super-Yang-Mills amplitudes, we now find in our constructed gravity theory non-vanishing four-point interactions such as, e.g.:
\begin{equation}
  \mathcal{A}_4(\chi^{\text{VA}}{}^{+}\chi^{\text{VA}}{}^{-}{}h^{++}\bartaua)= \mathcal{A}_4( \lambda^{+} \lambda^{-} A^{+}A^{-})\otimes\mathcal{A}_4(\pi \pi A^{+}A^{+})\,.
\end{equation}
This is quantitatively different than the amplitude arising from the double-copy with only the covariantized NLSM of \eqn{gaugeNLSMaction}.
\subsection{Double-copy consistency of massive Yang-Mills}
The structure of the hidden tower of corrections also lends itself to studying the double-copy consistency of massive gauge theory. By starting with the Proca Lagrangian,
\begin{equation}
\mathcal{L}=-\frac{1}{4}\text{Tr}(F^2)+\frac{1}{2}m^2\text{Tr}\left[A_\mu A^\mu\right]
\end{equation}
we can recover the starting point of this paper, \eqref{gaugeNLSMaction}, by performing a Stuckelberg transformation on the gauge fields,
\begin{equation}
A_\mu \rightarrow \frac{i}{g} U^\dagger D_\mu U
\end{equation}
where the $U\in SU(N)$ group elements are expressed in terms of the Cayley parameterization as in \eqref{gaugeNLSMaction}, and $\Lambda = m^2/g^2$. The color-dual nature of this action was previously studied in \cite{Momeni:2020vvr} with application to the double-copy construction of dRGT gravity amplitudes. As we have outlined throughout the text, many of the irrelevant operators that one could add to \eqref{gaugeNLSMaction}, included in the tables of \sect{sect:tables}, will be required for double-copy consistency in higher-multiplicity tree amplitudes.  

\section{A flavorful approach to a color-dual vector-pion theory}
\label{sect:radicalModel}
Here we demonstrate a more radical possibility for building a color-dual gauged pion theory.
A recent study by one of the authors \cite{Pavao:2022kog} demonstrated that $D$-dimensional gauge theory amplitudes can be decomposed into building blocks that preserve all the partial amplitude relations of the parent vector theory. This expansion, referred to as the reducible-amplitude block-decomposition, takes the following form for Yang-Mills,
\begin{equation}
A^{\text{YM}}_{(\sigma)} = \sum_{k=1}^{\lfloor |\sigma|/2\rfloor}\sum_{\rho \in S^{2|k}_{\sigma} } \epsilon_{(\rho)} \Delta^{(\rho)}_{(\sigma)}
\end{equation}
where $\rho = (a_1b_1)\dots(a_kb_k)$ is a list of paired legs that appear as external labels in the ordered list, $\sigma$, and $\epsilon_{(\rho)} = (\epsilon_{a_1}\epsilon_{b_1})\cdots (\epsilon_{a_k}\epsilon_{b_k})$ are dot products of $D$-dimensional polarizations. The sum starts at $k=1$ since all kinematic monomials for Yang-Mills at tree-level contain at least one factor of $\epsilon_{(ab)}$. The observation relevant for the following discussion is that $\Delta^{(\rho)}_{(\sigma)}$ are kinematic functions that independently satisfy BCJ relations with a flavor structure determined by $\rho$. Furthermore, the building blocks, $\Delta^{(\rho)}_{(\sigma)}$, can be consistently combined to produce pion amplitudes via the dimensional reduction of ref. \cite{Cheung:2017yef},
\begin{equation}\label{eq:YZdimRed}
\mathcal{E}^Z_a = \frac{1}{g f_\pi} (k^\mu_a,0,ik^\mu_a), \qquad  \mathcal{E}^{Y}_a  = (\vec{0},1,\vec{0}),
\end{equation}
where momenta $\mathcal{K}_a$ live in the $d$-dimensional scattering subspace:
\begin{equation}
\mathcal{K}_a = (k^\mu_a,0,0).
\end{equation}
Applying this dimensional reduction to pure gluon amplitudes in $D=2d+1$ dimensions will yield NLSM tree-level amplitudes when two particle $(ij)$ are projected along $Y$-directions, and the remaining particles take on $Z$-polarizations \cite{Cheung:2017yef}:
\begin{equation}\label{eq:flavorNLSM}
A^{\text{NLSM}}_{(12...n)} = A^{\text{YM}}(Z_1,...,Y_i,...,Y_j,...,Z_n) \equiv \Delta^{(ij)}_{(12...n)}\Big|_{\epsilon\rightarrow k}.
\end{equation}
These amplitudes can also be generated using Feynman rules from a Lagrangian that is perturbatively equivalent to NLSM given a particular weight of $Z$-particles \cite{Cheung2016prv}. Since they are equivalent to a special dimensional reduction from Yang-Mills \cite{Cheung:2017yef}, they can be expressed in terms of $\Delta^{(ij)}_{(12...n)}$ by considering how \eqn{eq:YZdimRed} acts on the block decomposition~\cite{Pavao:2022kog}.

Now let's return to the starting point of this paper. In pursuit of constructing amplitudes for DBIVA dynamically coupled to spacetime via the double-copy, the goal was to identify a double-copy consistent theory of NLSM pion interacting with gluons. While the construction of \eqn{eq:flavorNLSM} introduces different particle flavors that obscures the scalar-exchange symmetry of the NLSM Lagrangian, it does suggest an alternative approach to higher derivative model analyzed in previous sections. Namely, one could augment the flavored dimensional reduction in \eqn{eq:YZdimRed} by introducing an additional gluon polarization that is restricted to the $d$-dimensional subspace: 
\begin{equation}\label{eq:dimRedglue}
\mathcal{E}^A_a = (\epsilon^\mu_a,0,0).
\end{equation}
Truncating gluons from $D=2d+1$ down to $d$-dimensions provides a potential path towards consistently coupling the longitudinal pions with gluons; albeit in a way that violates exchange symmetry between the external pions. Take for example, the following four-scalar five-point amplitudes at pion mass dimension:
\begin{align}\label{eq:flavorPiC}
A^{\text{YM}}_5(1_Z, 2_Y, 3_Y, 4_{Z}, 5_A) &=
\frac{g}{f_{\pi}^2}\left(\frac{ s_{24} \kappa_1^{(5)}}{s_{15}}-\frac{ s_{13} \kappa_4^{(5)}}{s_{45}}+\kappa_{24}^{(5)}\right) \frac{s_{14}}{s_{23}}\sim s_{14} \Delta^{(14)(23)}_{(12345)},
\\\label{eq:flavorPiA}
A^{\text{YM}}_5(1_Y, 2_Z, 3_Z, 4_{Y}, 5_A) &=
\frac{g}{f_{\pi}^2}\left(\frac{ s_{24} \kappa_1^{(5)}}{s_{15}}-\frac{ s_{13} \kappa_4^{(5)}}{s_{45}}+\kappa_{24}^{(5)}\right) \sim s_{23} \Delta^{(14)(23)}_{(12345)},
\\ \label{eq:flavorPiB}
A^{\text{YM}}_5(1_Y, 2_Z, 3_{Y}, 4_Z, 5_A) &=
\frac{g}{f_{\pi}^2} \left(\frac{ s_{24} \kappa_1^{(5)}}{s_{15}}-\frac{ s_{24}\kappa_4^{(5)}}{s_{45}}\right) \sim s_{24} \Delta^{(13)(24)}_{(12345)}.
\end{align}
where the relationship of these amplitudes to $\Delta^{(\rho)}_{(\sigma)}$ is due to the Ward identities derived in \cite{Pavao:2022kog}. It is worth noting that $A^{\text{YM}}_5(1_Y, 2_Z, 3_Z, 4_{Y}, 5_A)$ happens to be identical to the last term in \eqn{eq:dcc5point}. We remind the reader that this term is precisely the five-point amplitude,  $A_5^{\text{cov.}\pi}$, generated by the gauged NLSM Lagrangian of \eqn{gaugeNLSMaction}:
\begin{equation}
A_5^{\text{cov.}\pi}(1_\pi, 2_\pi, 3_\pi, 4_\pi, 5_A) \equiv A^{\text{YM}}_5(1_Y, 2_Z, 3_Z, 4_{Y}, 5_A).
\end{equation}
However, given this definition, one can check that the four-pion five-point amplitude of $A^{\text{cov}.\pi}_5$ does not satisfy BCJ relations when the pions are taken to be \textit{identical bosons}. This observation led earlier to introducing an additional $\text{Tr}(F^3)$ operator with coupling $\Lambda/g$ in \sect{sect:trF3Intro}. Indeed, one can check that the following BCJ relation, \eqn{eq:BCJrel},  does not hold when applied to $A_5^{\text{cov.}\pi}$:
\begin{equation}
\begin{split}
s_{1|2}A_5^{\text{cov.}\pi}(1_\pi, 2_\pi, 3_\pi, 4_\pi, 5_A) &+ s_{13|2}A^{\text{cov.}\pi}_5(1_\pi,  3_\pi,2_\pi,  4_\pi, 5_A) 
\\
&+ s_{134|2}A^{\text{cov.}\pi}_5(1_\pi,  3_\pi, 4_\pi, 2_\pi,5_A) \neq 0,
\end{split}
\end{equation}
where $s_{i|j_1...j_n} = k_i\cdot(k_{j_1}+\cdots +k_{j_n})$. 

In contrast, when we keep track of \textit{particle flavor} for the dimensionally reduced pionic amplitudes of \eqn{eq:flavorPiA} and \eqn{eq:flavorPiB}, with $Y$- and $Z$-particles coupled to gluons, we find that the usual BCJ relation does hold:
\begin{equation}
\begin{split}
s_{1|2}A^{\text{YM}}_5(1_Y, 2_Z, 3_Z, 4_{Y}, 5_A) &+ s_{13|2}A^{\text{YM}}_5(1_Y, 3_Z,2_Z,  4_{Y},  5_A) 
\\
&+ s_{134|2}A^{\text{YM}}_5(1_Y, 3_Z, 4_{Y},2_Z,  5_A) =0.
\end{split}
\end{equation}
This is simply a consequence of Yang-Mills satisfying BCJ relations in $D$-dimensions, and the pion-vector amplitudes above just being a special choice of $D$-dimensional polarizations. Thus, the amplitudes generated using the dimensional reduction of \eqn{eq:YZdimRed} and \eqn{eq:dimRedglue} are manifestly color-dual.  Indeed an important feature of this model is that any $m$-point tree-level amplitude with $2(m-1)$ $Z$ particles and 2 $Y$ particles will match exactly the same scattering amplitudes and have the same soft behavior as the NLSM.

Since we require no additional interactions beyond those generated by the dimensional reduction from $D=2d+1$ dimensions, the six-point pure-scalar NLSM amplitude is unaffected, and consequently the behavior of six-point DBIVA amplitudes in the double-copy theory coupled to supergravity is likewise unaffected.  It is, however, worth noting  that \eqn{eq:flavorPiC} has additional poles  not generated by the gauged NLSM of \eqn{gaugeNLSMaction}. In this case, unlike the approach studied in previous sections, these additional residues avoid the introduction of $\text{tr}(F^3)$ that forced a tower of higher-derivative pion-vector interactions. These additional poles are generated by a combination of the pure-scalar three-point vertices derived in \cite{Cheung:2017yef}, and additional vector-pion interactions. The following are the off-shell Feynman rules for the novel interactions between gluons and $Y$-particles,
\begin{equation}
\label{eq:YYAVertex}
\threeGraph{scalar}{scalar}{gluon}{Y_1^a}{Y_2^b}{A_3^{c,\mu}}\equiv V_{abc}^{\mu}=g\, f^{abc}(k_1-k_2)^\mu ,
\end{equation}
and gluons with the $Z$-particle,
\be\label{eq:AAZVertex}
\threeGraph{gluon}{gluon}{}{A_1^{a,\mu}}{A_2^{b,\nu}}{Z_3^c}\equiv V_{abc}^{\mu\nu}= f_{\pi}^{-1}\,f^{abc}\big[\eta^{\mu\nu}\,k_3\!\cdot\!(k_1-k_2) +(k_3^\nu k_2^\mu-k_3^\mu k_1^\nu)\big].
\\ \
\ee
Both of these vertex rules can be constructed by applying the dimensional reduction of \eqn{eq:YZdimRed} to the off-shell Feynman rules of the Yang-Mills three-point function in \eqn{fSqNumThree}. While the vertex $V_{abc}^{\mu}$ of \eqn{eq:YYAVertex} produces the familiar three-point amplitude of \eqn{covScalarThree}, the vertex $V_{abc}^{\mu\nu}$ of \eqn{eq:AAZVertex} is novel.  A consequence is that any tree-level amplitudes involving external $Y$ and $A$ external states will simply be amplitudes shared with the covariantized free scalar. 

Furthermore, while $V_{abc}^{\mu\nu}$ is needed for consistent factorization of the $A^{\text{YM}}_{(AZZA)}$ four-point amplitude, we can see that it must vanish on-shell\footnote{The $A^{\text{YM}}_{(AAZ)}$ amplitude is equivalent on-shell to three-point function $A^{(\chi)}_{(gg\pi)}$ of \cite{Pavao:2022kog}, in that they both vanish. However as we note, $A^{\text{YM}}_{(AAZ)}$ vanishes due to on-shell three-point kinematics, whereas the three-point function $A^{\chi}_{(gg\pi)}$ vanished due to the kinematic factor being symmetric under color ordering.} when external polarizations are transverse and momenta are light-like. In addition to $V_{abc}^{\mu\nu}$ at three-point, there will also be novel four-point vertices that are required for  $A^{\text{YM}}_{(AZZA)}$ to be gauge invariant. 

In summary, while this approach violates exchange symmetry between external scalars, it has merit in that we can smoothly interpolate between pure Yang-Mills amplitudes, when taking $f_\pi \rightarrow \infty$, and pure NLSM amplitudes, when taking the gauge coupling $g\rightarrow 0$. Furthermore, since this flavored formulation of gauged pions avoids the $\mathcal{O}(\Lambda^2/g)$ five-point contribution described in \sect{sect:NLSMmod}, the Adler's zero of pure pions at six-point is preserved. In other words, any pure scalar amplitudes at pion mass dimension are identical to those generated by the NLSM Lagrangian, free of corrections from higher-dimension contacts needed for double-copy consistency.

%
%
 \section{Conclusions}\label{sect:Conclusions}
In this paper we have begun to establish whether a color-dual theory of vectors coupled to pions can exist.  We have demonstrated the existence of two distinct candidate theories through six-point interactions. We include expressions for all basis kinematic weights $n(g)$ and associated amplitudes in a machine readable auxiliary file.  One fascinating promise of such a color-dual theory is its role in the double-copy description of a Dirac-Born-Infeld-Volkov-Akulov-type theory coupled to dynamic spacetime -- a necessary ingredient if we aspire to double-copy descriptions of many phenomenological $\alpha$-attractor models involving supergravity.  As the literature provides color-dual maximally supersymmetric Yang-Mills theory representations through four-loops~\cite{Bern:2012uf}, we have reduced multi-loop calculations in $\text{DBIVA}+\text{CSG}$ to carrying out cut-construction in our candidate color-dual vector-pion theory. 

The possibilities for future study are rather large, with non-trivial payoff, and these are just first steps. To craft arbitrary inflationary potentials we will want to investigate double-copy construction of higher-derivative operators targeting the inflaton sector.  It will be intriguing to explore the consequences of these choices especially considering the good UV behavior of conformal supergravity theories, which seem a necessary feature of our more conservative double-copy construction of \sec{sect:bootstrapF3}. This model, however, suffers from modified Adler's zero behavior which could call into question whether the fermions in the double-copy supergravity theory can rightly be associated with nilpotent superfields -- a question that awaits future study. Our second model described in \sec{sect:radicalModel} neatly evades this issue at the cost of manifest exchange symmetry between scalars in the color-dual gauged pion theory.

Towards constructing the type of ultra-soft UV behavior potentially offered by string theory one should, in both models, consider questions of attempting to climb the ladder of higher-derivative operators towards closed-string scattering with the abelianized open-string.  Furthermore, it will be fascinating to see what type of interpretation one can give to other double-copy constructions associated with these color-dual vector-pion theories and to explore their potential phenomenology.  Consider, for example, the double-copy of each color-dual gauged-pion model  with itself. By identifying color-dual gauged pion theories, we have effectively constructed color dual theories describing a decoupling limit of massive vectors -- theories whose consistent double-copy should describe a decoupling limit of massive gravity. Indeed, the resulting double-copy spectrum should describe Goldstone modes of broken time-translation generators in a theory of $\mathcal{N}=0$ gravity + Born-Infeld + Galileons, arguably a theory of dark energy.

\paragraph{Acknowledgements:}
We  gratefully acknowledge Alex Edison, Renata Kallosh, Ian Low, James Mangan, Radu Roiban, Aslan Seifi and Suna Zekio\u{g}lu for related collaboration, useful conversations, feedback on earlier drafts, and encouragement at various stages. This work was supported by the DOE under contracts  DE-SC0021485 and DE-SC0015910, and by the Alfred P. Sloan Foundation. 
\appendix

\bibliographystyle{JHEP}
\bibliography{Refs_factoringInflation}
\end{document}